\documentclass[superscriptaddress,twocolumn]{revtex4-2}

\usepackage{graphicx}% Include figure files
\usepackage{placeins}% To stop the float
\usepackage{dcolumn}% Align table columns on decimal point
\usepackage{bm}% bold math
\usepackage[colorlinks=true, citecolor=cyan, urlcolor = violet, linkcolor= purple, bookmarks=true]{hyperref}
\usepackage{float}
\usepackage{amsmath,amsfonts,amssymb}
\usepackage{natbib}
\usepackage{hyperref}
\usepackage{breakurl}
\usepackage{xcolor,verbatim,multirow}
\usepackage[normalem]{ulem}
\usepackage{subcaption} 
\usepackage{wrapfig}
\usepackage{adjustbox}
\usepackage{caption}
\begin{document}
%----------------------------------------------------------------------
%Title
%----------------------------------------------------------------------
\title{Investigating effects of dark matter on photon orbits and black hole shadows}
%\title{Signatures of dark matter in the EHT images of M87* and Sgr A*}
%----------------------------------------------------------------------
%Authors
%----------------------------------------------------------------------
\author{Arshia Anjum}
\email{anjumarshia271@gmail.com}
\affiliation{School of Physical Sciences, National Institute of Science Education and Research, HBNI, Jatni-752050, Orissa, India}
\author{Misba Afrin}
\email{me.misba@gmail.com}
\affiliation{Centre for Theoretical Physics,
Jamia Millia Islamia, New Delhi 110025, India}
\author{Sushant G. Ghosh} \email{sghosh2@jmi.ac.in, sgghosh@gmail.com}
\affiliation{Centre for Theoretical Physics, Jamia Millia Islamia, New Delhi 110025, India}
\affiliation{Astrophysics and Cosmology	Research Unit, School of Mathematics, Statistics and Computer Science, University of KwaZulu-Natal, Private Bag 54001, Durban 4000, South Africa}
%----------------------------------------------------------------------
%Abstract
%----------------------------------------------------------------------
\begin{abstract}
We consider Kerr black holes (BHs) surrounded by perfect dark fluid matter (PFDM), with an additional parameter ($k$) because of PFDM, apart from mass ($M$) and rotation parameter ($a$) --- the rotating PFDM BHs. We analyse the photon orbits around PFDM BHs and naked singularities (NSs) and emphasize the effect of PFDM on photon \emph{boomerangs}. Interestingly, the azimuthal oscillations first increase and then decrease for retrograde orbits, whereas they first decrease and then increase for prograde orbits, with increasing $k$. Unlike in the Kerr NSs, photon \emph{boomerangs} can form around rotating PFDM NSs. We use the Event Horizon Telescope (EHT) observational results for Schwarzschild shadow deviations of M87* and Sgr A*, $\delta_{M87^*}=-0.01\pm0.17$ and $\delta_{Sgr A^*} = -0.08^{+0.09}_{-0.09}~\text{(VLTI)},-0.04^{+0.09}_{-0.10}~\text{(Keck)}$, to report the upper bounds on the PFDM parameter: $0\leq k\leq 0.0792M$ and $k^{max}\in[0.0507M, 0.0611M]$ respectively. Together with the EHT bounds on the shadows of Sgr A$^*$ and M87$^*$, our analysis concludes that a substantial part of the rotating PFDM BH parameter space agrees with the EHT observations. Thus, one must consider the possibility of the rotating PFDM BHs being strong candidates for the astrophysical BHs.
\end{abstract}
%\keywords{BHs, NSs, generating solution, Vaidya solution, Type II fluid}
                              
\maketitle
%----------------------------------------------------------------------
%Body of Document
%----------------------------------------------------------------------
\section{Introduction}\label{intro}
Black holes (BHs), one of the most exceptional predictions of the theory of General Relativity (GR), provide a means to probe theories of gravity through multifarious strong field phenomena like the formation of black hole shadows \citep{Bardeen:1973tla}. Based on the uniqueness theorem, the Kerr hypothesis states that astrophysical BHs are uniquely described by the Kerr metric \citep{Kerr:1963ud} -- the only stationary, axially symmetric, and asymptotically flat vacuum solution of the Einstein equations \citep{Carter:1968rr}. 
The first ever horizon-scale images of the M87* \citep{EventHorizonTelescope:2019dse,EventHorizonTelescope:2019pgp} and Sgr A* \citep{EventHorizonTelescope:2022xnr,EventHorizonTelescope:2022xqj} BHs by the Event Horizon Telescope (EHT) indicate the consistency of the observed shadows with that expected for a Kerr BH in GR. Hence, studying photon motion around the Kerr spacetime, is exceptionally relevant to the current observations. A particular class of photon orbits with constant radii, the spherical photon orbits, which trace out the shadow silhouette, is astrophysically relevant among the photon orbits around a rotating spacetime \citep{Wilkins:1972rs}. The exact solutions for radii in this class of light orbits, in terms of the BH spin parameter and the trajectory's effective inclination angle, have been obtained for two particular cases, viz. equatorial orbits (called light rings) and the polar orbit \citep{Chandrasekhar:1985kt,Teo:2020sey}. However, we are interested in the spherical photon orbits between these two extremes, i.e., photons moving along the constant radial coordinate trajectories whose motion constants govern the local escape cones of photons related to any family of observers, and especially the shadow of BHs located in front of a radiating source. The spherical timelike orbits around the Kerr BH was proposed by Wilkins \citep{Johnston:1974pn} and extended to the Kerr-Newman case \citep{Johnston:1974pn}. 
\begin{figure}[t]
\centering
\includegraphics[width = 0.45\textwidth]{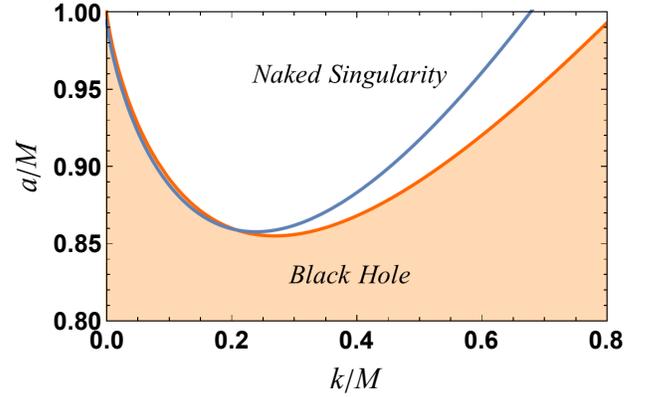}
\caption{Parameter space of rotating PFDM BHs.  The red line separates BHs from NSs ($\Delta(r)=0$ has no real roots). The extremal rotating PFDM BH occurs for $k_{E1}=0.086M$, $k_{E2}=0.525M$ at $a=0.9M$ and for $a_E=0.525M$ at $k=0.6M$. The blue line corresponds to allowed ($k$, $a$) for photon \emph{boomerang}. Clearly, photon \emph{boomerang} can occur around both BHs and NSs for $k\neq0$ (see Section~\ref{sect:Visualization}, for details on \emph{boomerang}).} \label{fig:parameterSpace}
\end{figure}
\begin{figure*}[t]
\begin{center}
    \begin{tabular}{c c}
    \includegraphics[width=0.45\textwidth]{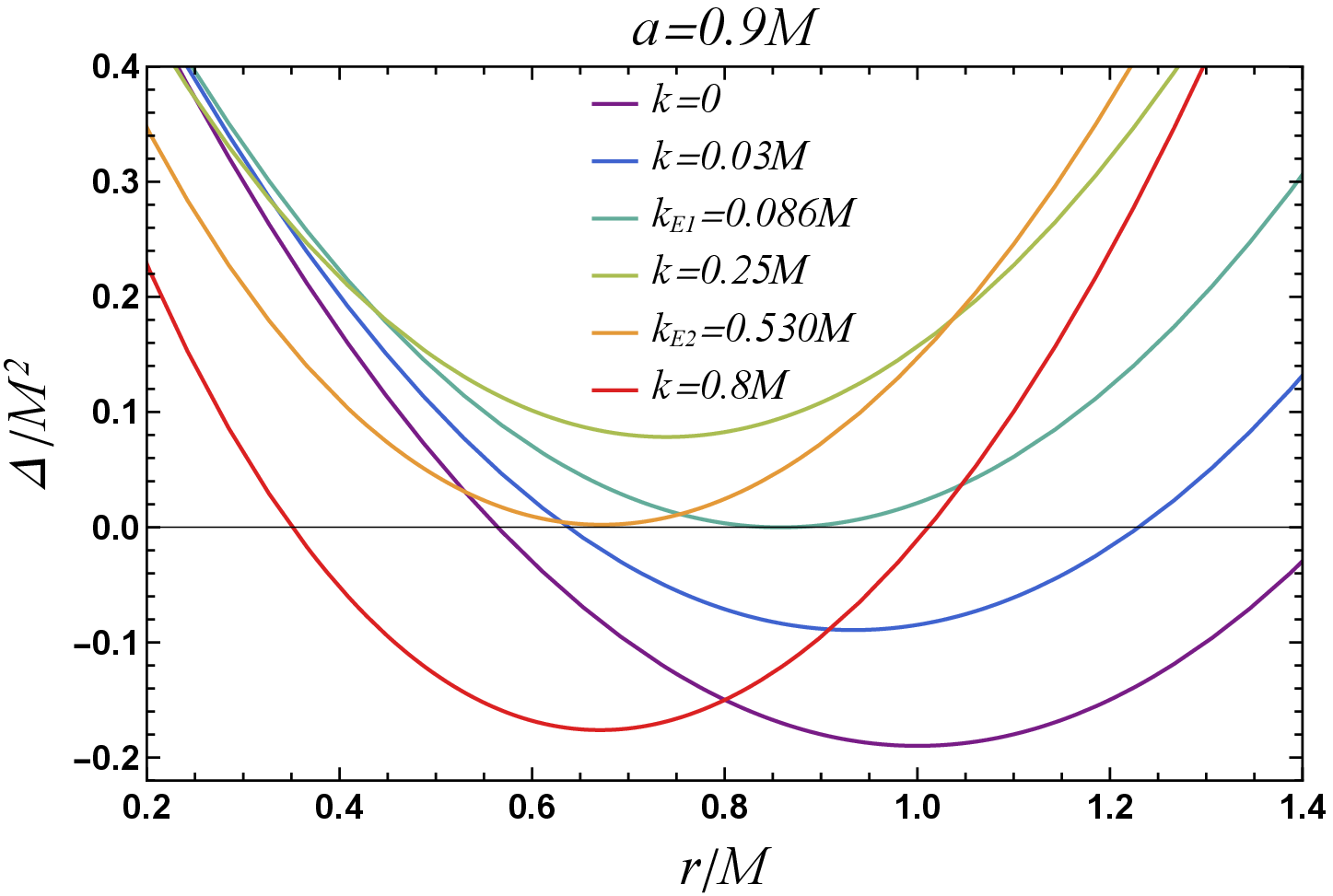}&
     \includegraphics[width=0.45\textwidth]{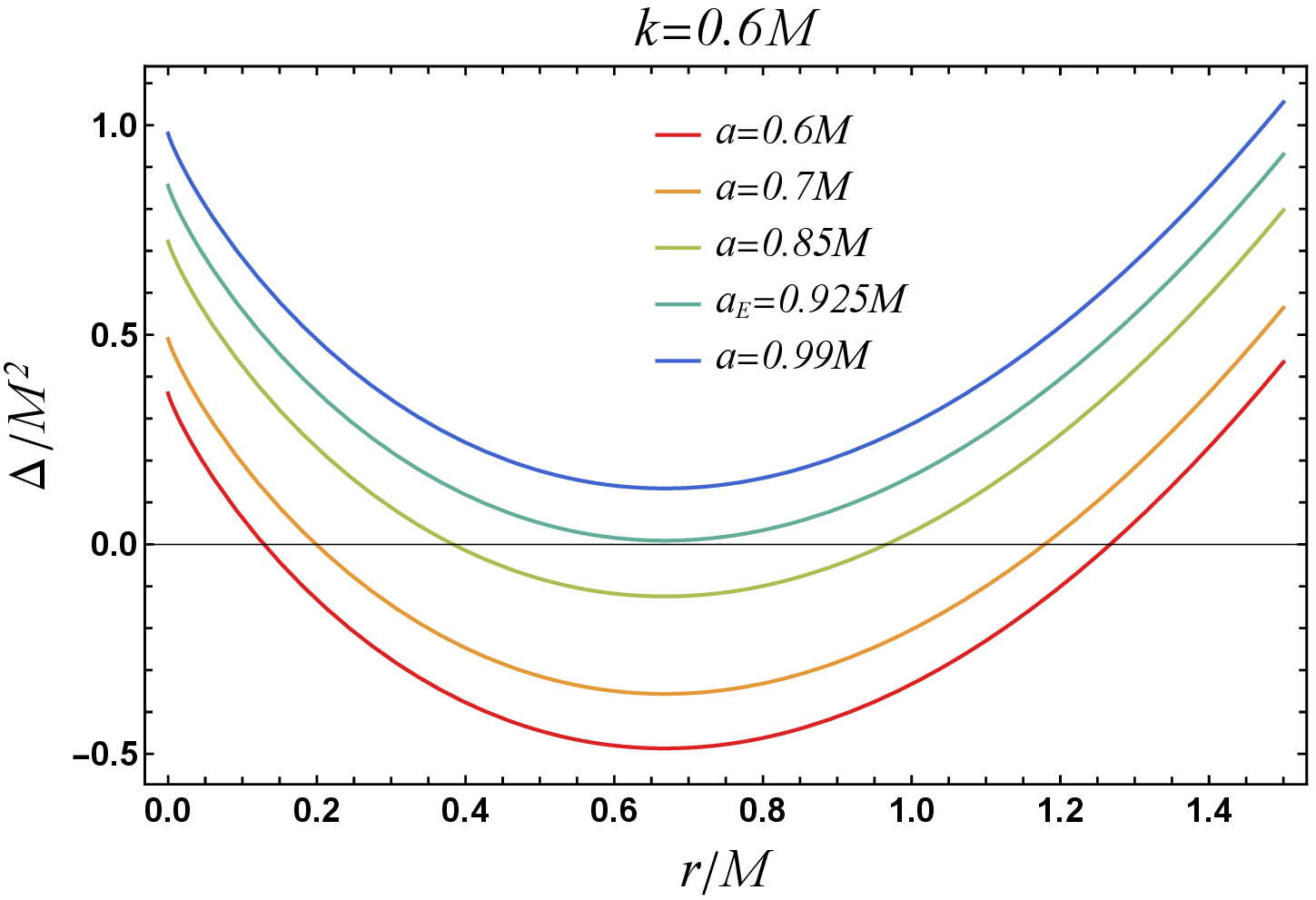}
    \end{tabular}
\end{center}
\caption{Horizons of the rotating PFDM BHs. For a given $a\, (\gtrsim 0.855M)$, there are two extreme values of PFDM parameter viz. $k=k_{E1},\, k_{E2}$, such that $\Delta=0$ admits a double root corresponding to an extremal BH with degenerate horizons, whereas when $k_{E1}<k<k_{E2}$, we have an NS and $k<k_{E1}$ or $k>k_{E2}$ correspond to BHs with Cauchy and event horizons (left). However, for a given $k$, there is only one extreme value $a=a_E$, such that $\Delta=0$ has a double root corresponding to an extremal black hole (right). }
\label{fig:HorizonPlot}
\end{figure*}
Later, the properties of spherical photon orbits outside the Kerr BH's outer horizon has been numerically  \citep{Teo:2020sey}. It has been extended to analytical solutions by showing two exterior non-equatorial photon orbits \citep{Tavlayan:2020cso}. Noteworthy attention is devoted to generalize the spherical photon orbits around Kerr BHs, for instance, to the case for Kerr naked singularities (NSs) \citep{Charbulak:2018wzb}, near-extremal Kerr BHs \citep{Igata:2019pgb}, deformed Kerr BH with an extra deformation parameter\citep{Liu:2017ifc}, and Myers–Perry BH \citep{Bugden:2018uya}. Among the spherical photon orbits, an interesting case of photons returning to the same point in exactly opposite direction after one complete azimuthal oscillation -- the photon \emph{boomerang}-- has been studied around the Kerr BH \citep{Page:2021rhx}.

Remarkably, the BHs bear imprints of not only the intrinsic spacetime geometry, but of exotic surrounding matter as well, and thus give a clue about different matter-energy distributions in its immediate environment \citep{Chen:2005qh,Lacroix:2012nz, Ghosh:2015ovj,Ghosh:2014pga,Ghosh:2015ovj,Ghosh:2016ddh,Ma:2020dhv,Saurabh:2020zqg,Das:2020yxw,Atamurotov:2021hoq,Atamurotov:2021hck}. Furthermore, the BH shadow is a tool to constrain the otherwise obscure surrounding matter distribution \citep{DePaolis:2010vw,Kumar:2017tdw,Jusufi:2020zln,Nampalliwar:2021tyz,Pantig:2021zqe,Vagnozzi:2022moj,Pantig:2022whj} which include, beside luminous matter, signatures of dark matter. The dark matter is one of the open fundamental questions in physics, and it makes up 25\% of the Universe's energy density \citep{ParticleDataGroup:2018ovx}. Despite conflicting with long-standing (and more recent) small-scale structure evidence, the cold dark matter model is now the most popular dark matter model. The phenomenological PFDM model \citep{Hou:2018avu, Haroon:2018ryd,Das:2020yxw,Das:2021otl,Narzilloev:2020qtd,Atamurotov:2021hoq,Atamurotov:2021hck}, in which dark matter is depicted as a perfect fluid, is another possible dark matter explanation.
We shall try to extract information about dark matter by investigating rotating spacetimes surrounded by PFDM \citep{Hou:2018avu}. We intend to analyse the effect of the PFDM on the spherical photon orbits around BHs and NSs. We will extend the previous analysis in \citep{Page:2021rhx} to find the possible effects of surrounding PFDM on the \emph{boomerang} orbits and also to inquire whether such orbits can be formed around NSs as well. Further, we will investigate how the PFDM affects the black hole shadow structure and compare our result with those of the Kerr BHs' \citep{Bugden:2018uya}. The Kerr BHs surrounded with PFDM, which henceforth, for definiteness, will be addressed as rotating PFDM BHs, may arise because of surrounding fluid-like dark matter. Further, we will inquire whether the EHT results of M87* and Sgr A* could shed light on dark matter's properties \citep{Hou:2018avu,Haroon:2019new,Xu:2018mkl,Atamurotov:2021hoq,Atamurotov:2021hck} and in turn constrain the amount of PFDM. The EHT has inferred mass and distance of both M87* \citep{EventHorizonTelescope:2019dse,EventHorizonTelescope:2019pgp} and Sgr A* \citep{EventHorizonTelescope:2022xnr,EventHorizonTelescope:2022xqj} and put bounds on the dimensions of the central brightness depression by calibrating the ring image to the shadow size in case of M87* and by direct measurement of the shadow size in case of Sgr A*, to obtain the fractional deviation between the model shadow size and the Schwarzschild shadow size $\delta_{M87^*}=-0.01\pm0.17$ \citep{EventHorizonTelescope:2019ggy,EventHorizonTelescope:2021dqv} and $\delta_{Sgr A^*}=-0.08^{+0.09}_{-0.09}~\text{(VLTI)},-0.04^{+0.09}_{-0.10}~\text{(Keck)}$ \citep{EventHorizonTelescope:2022xnr,EventHorizonTelescope:2022xqj}. Modelling the rotating PFDM BHs as M87* and Sgr A*, we inquire whether they can be candidates for supermassive BHs and also put astrophysical limits on the amount of dark matter, directly, via the BH shadows.

We organize the paper as follows: in Section \ref{sect:EquationsofMotion}, we present the rotating PFDM metric, obtain the photon geodesics and discuss the effect of PFDM on the horizon structure as well as on the latitudinal motion of spherical photons and the photon region. The Section \ref{sect:Visualization} pertains to the visualization of the spherical photon orbits around rotating PFDM BHs and NSs and a special case of photon \emph{boomerang} is also studied. The Section \ref{sect:constraints} is devoted to the construction of shadow silhouettes and constraining $k$ from the EHT-inferred bounds on the Schwarzschild shadow deviation $\delta$ of M87* and Sgr A*. Finally, concluding remarks are drawn in section \ref{Conclusion}.

We use geometrized units $8 \pi G= c =1$, unless the units are specifically defined.
\begin{figure*}[t]
\begin{center}
    \begin{tabular}{c c}
    \includegraphics[width=0.4\textwidth]{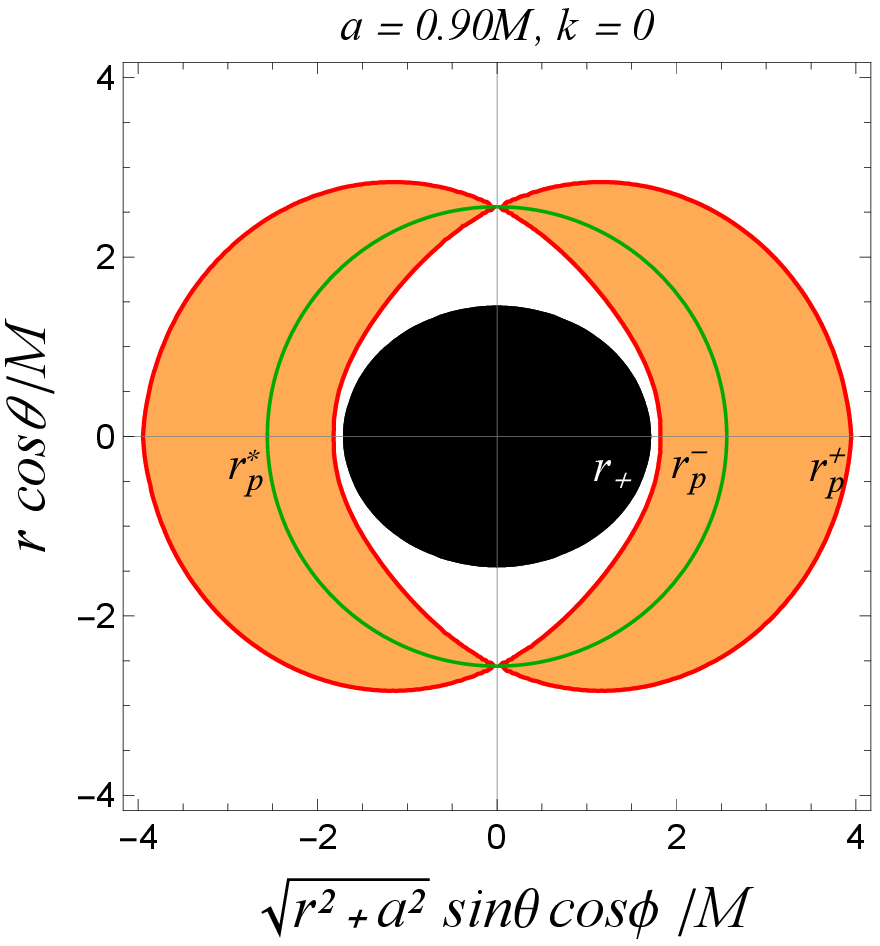}&
     \includegraphics[width=0.4\textwidth]{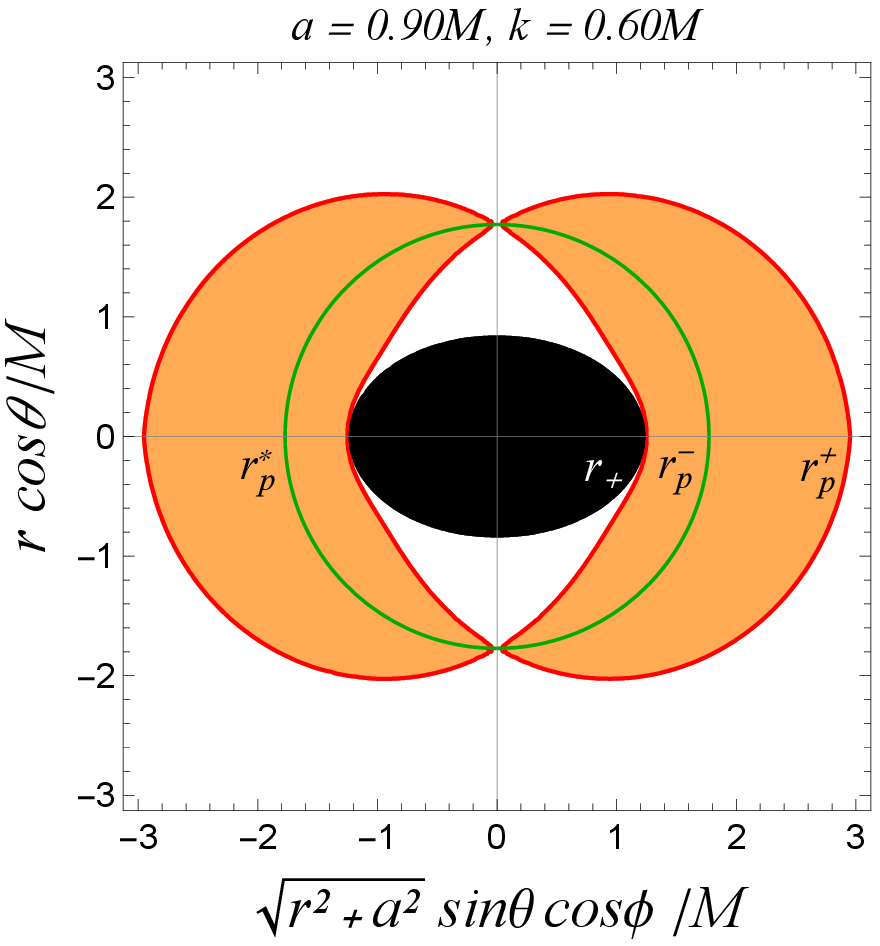}
    \end{tabular}
\end{center}
\caption{The photon region (orange) around a Kerr BH (left) and rotating PFDM BH (right). The boundary of the central black region corresponds to an event horizon. (For interpretation of the references to colour in this figure legend, the reader is referred to the web version of this article).} \label{fig:PhotonRegion}
\end{figure*}
%%%%%%%%%%%%%%%%%%%%%%%%%%%%%%%%%%%%%%%%%%%%%%%%%
\section{Equations of motion in rotating PFDM spacetime} \label{sect:EquationsofMotion}
The action corresponding to the dark matter field minimally coupled to gravity is  given by \citep{Xu:2017bpz}
\begin{equation}
    S=\int d^4x \sqrt{-g}\,\left(\frac{R}{16\pi}+\mathcal{L}_{DM}\right),\label{action}
\end{equation}
where $\mathcal{L}_{DM}$ is the dark matter Lagrangian density, which is considered to be non-interacting with ordinary matter. Consequently, the Einstein field equations become \citep{Li:2012zx,Xu:2017bpz,Das:2021otl}
\begin{equation}
    R_{\mu\nu}-\frac{1}{2}g_{\mu\nu}R=8\pi(T^{M}_{\mu\nu}-T^{DM}_{\mu\nu})=8\pi T_{\mu\nu},\label{field_eq}
\end{equation}
where $T^{M}_{\mu\nu}$ and $T^{DM}_{\mu\nu}$ are the energy-momentum tensor of the ordinary matter and dark matter respectively. Considering the surrounding dark matter to be a perfect fluid the energy momentum tensor reads \citep{Xu:2017bpz}
\begin{equation}
   T_{\mu}^{\nu\,DM}=\text{diag}[-\rho,p_r,p_\theta,p_\phi],
\end{equation}
where $\rho$ is the energy density of dark matter and $p_r$, $p_\theta$, $p_\phi$ are the pressures along different directions. In the simplest case, the equation of state of the perfect fluid dark matter (PFDM) can be assumed to be $p_\theta=p_\phi=(\delta-1)\rho$ where $\delta$ is a constant, and we are interested in the case $\delta=3/2$ \citep{Li:2012zx}. Thus, the spherically symmetric solution with surrounding PFDM reads \cite{Hou:2018avu}
\begin{align} \label{metric_spherical}
    ds^2&=-\left[1-\frac{2m(r)}{r}\right]dt^2+\left[1-\frac{2m(r)}{r}\right]^{-1}dr^2\\ \nonumber
    &+r^2(d\theta^2+\sin^2{\theta}d\phi^2),
\end{align}
with the black hole mass $M$ being modified to \cite{Hou:2018avu}
\begin{equation}
    m(r)= M-\frac{k}{2}\ln{\left(\frac{r}{|k|}\right)},\label{mass_function}
\end{equation}
where $k$ is an integration constant which quantifies the amount or intensity of the PFDM. The energy density and pressures are given by \cite{Das:2021otl}
\begin{equation}
    \rho=-p_r=\frac{k}{8\pi r^3}\,\,\, \text{and}\,\,\, p_\theta=p_\phi=-\frac{k}{8\pi r^3}.
\end{equation}
To satisfy weak energy conditions, we must have $\rho\geq 0$, which necessitates $k\geq 0$, which is hence assumed throughout this work.

The rotating counterpart of (\ref{metric_spherical}), obtained using a modified Newman-Janis algorithm \citep{Azreg-Ainou:2014pra}, can be casted in a Kerr-like form as \citep{Hou:2018avu,Xu:2017bpz,Das:2021otl}
\begin{eqnarray}\label{metric}
ds^2 & = & -\left(1-\frac{2m(r)r}{\Sigma} \right) dt^2  - \frac{4am(r)r}{\Sigma} \sin^2 \theta dt\:d\phi + \frac{\Sigma}{\Delta}dr^2  \nonumber\\
 & &+ \Sigma d \theta^2+ \left[r^2+ a^2 + \frac{2m(r) r a^2 }{\Sigma} \sin^2 \theta
\right] \sin^2 \theta d\phi^2,
\end{eqnarray}
%%%%%%%%%%%%%%%%%%%%%%%%%%%%%%%%%%%%%%%%%%%%%%%%%%%
\begin{figure*}[t]
\begin{center}
    \begin{tabular}{c c}
    \includegraphics[width=0.49\textwidth]{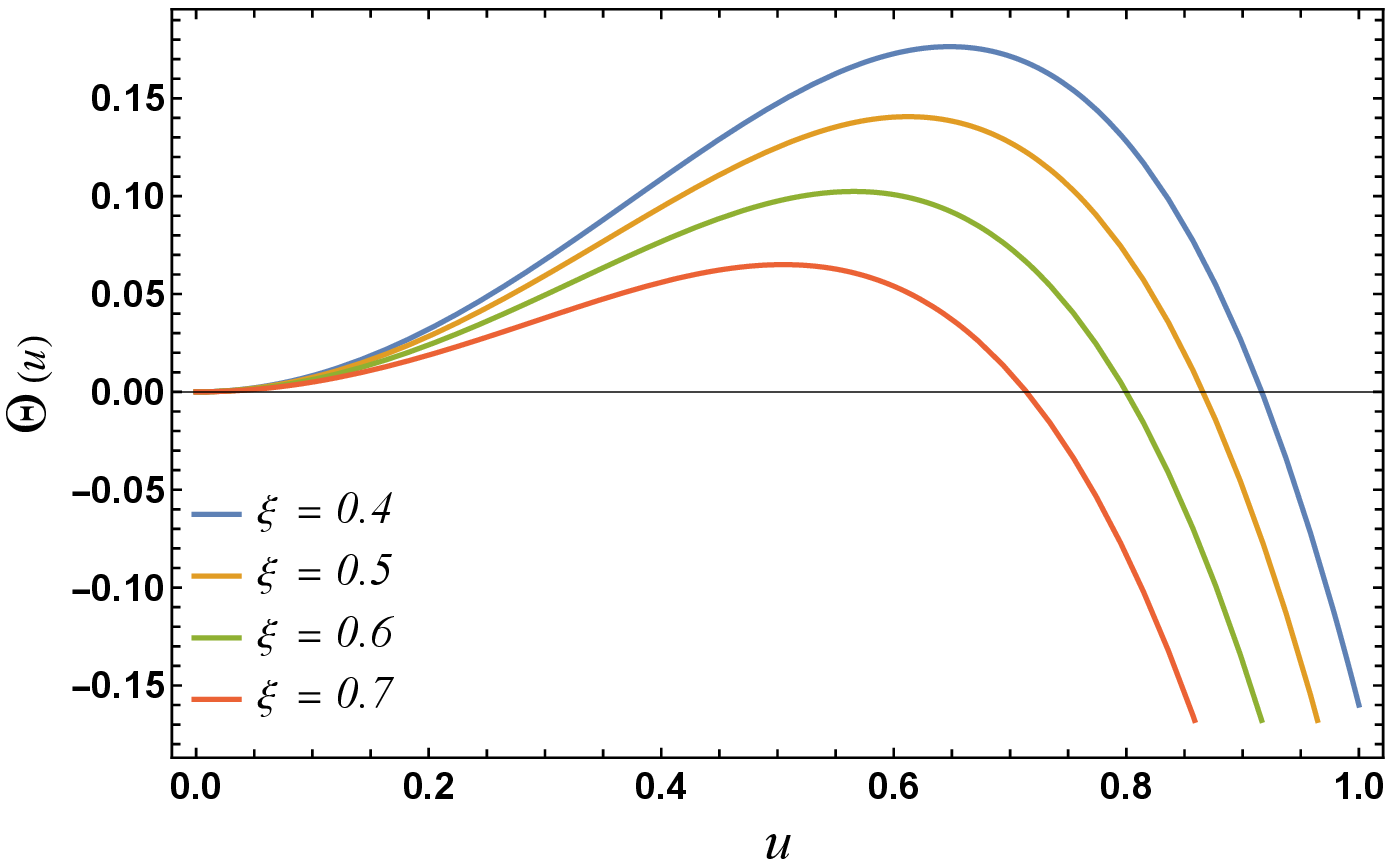}&
     \includegraphics[width=0.49\textwidth]{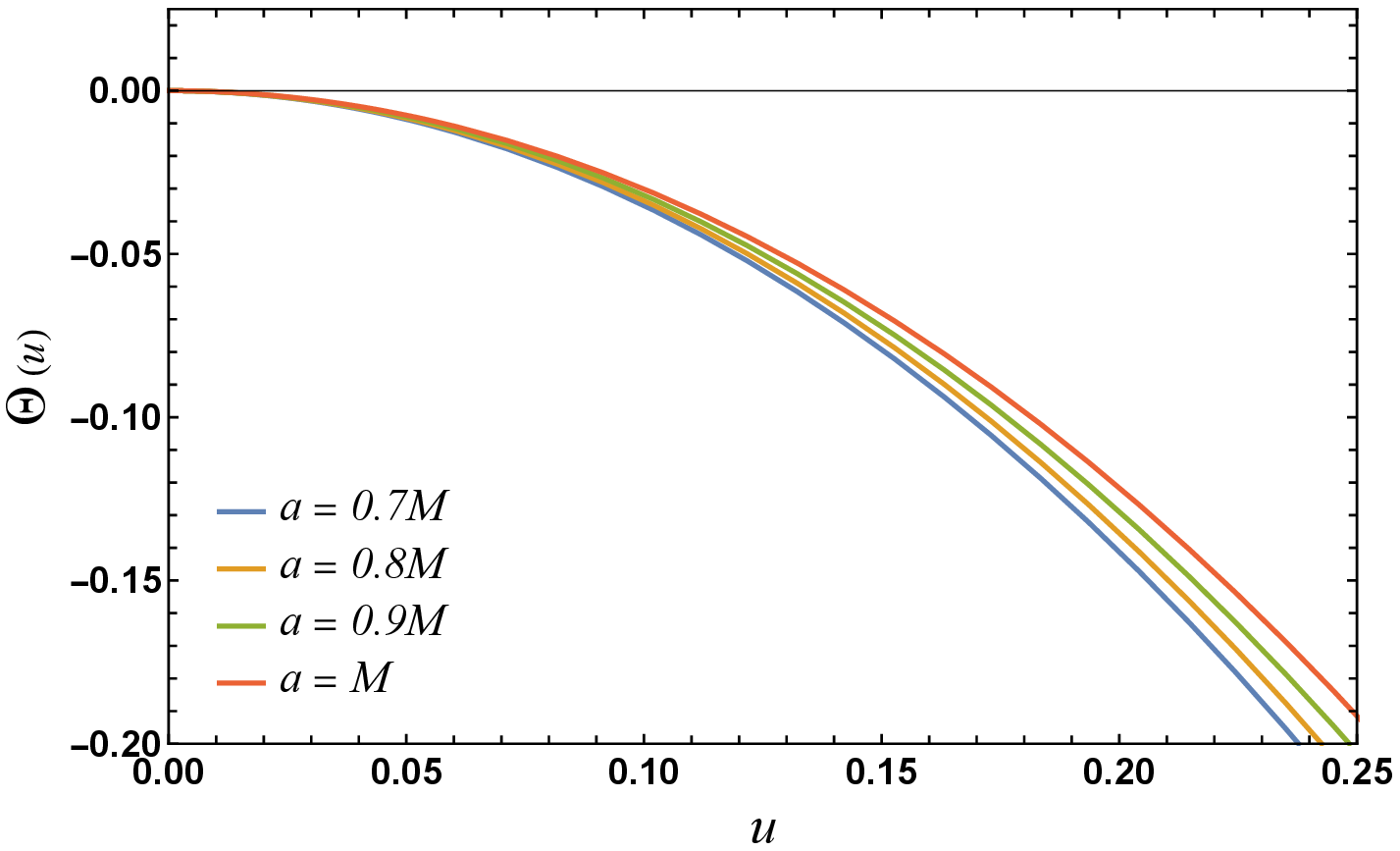}
\end{tabular}
\end{center}
	\caption{Plot showing the behaviour of $\Theta(u)$ ($\eta = 0$), such that, (a) $\xi^2 < a^2 $ (with $a\approx M$) (left) and (b) $\xi^2 \geq a^2$ (with $\xi=2$) (right).}
	\label{fig:u_eqn_1}
\end{figure*}
%%%%%%%%%%%%%%%%%%%%%%%%%%%%%%%%%%%%%%%%%%%%%%%%%%%%%
where $\Sigma = r^2 + a^2 \cos^2\theta$ and  $\Delta = r^2 + a^2 - 2m(r)r$.
Here is $a$ the spin parameter and the metric (\ref{metric}) go over to Kerr spacetime in absence of the surrounding PFDM ($k=0$). In general, the metric (\ref{metric}) is a prototype non-Kerr spacetime that represents Kerr-Newman \citep{Newman:1965my} and Kerr \citep{Kerr:1963ud} spacetimes, respectively, when $m(r)=M-{Q^2}/{2r}$ and $m(r)=M$; further, depending on $m(r)$, the metric (\ref{metric}) can also represent a variety of BHs viz., Hayward \citep{Kumar:2020yem,KumarWalia:2022aop}, Bardeen \citep{Kumar:2020yem,KumarWalia:2022aop}, Ghosh-Culetu \citep{Kumar:2020yem,KumarWalia:2022aop}, Ghosh-Kumar \citep{KumarWalia:2022aop}, rotating Quintessence \citep{Ghosh:2015ovj}, hairy Kerr \citep{Afrin:2021imp,Islam:2021dyk}, rotating Horndeski \citep{Afrin:2021wlj}.
Indeed, dark matter makes up a major portion of the Universe's energy density \citep{ParticleDataGroup:2018ovx} and any observational constraints placed on $k$ would be an essential step in fundamental physics. The  null surface $\Sigma\neq0,\Delta=0$ corresponds to a coordinate singularity. Solving $\Delta=0$ leads to two positive real roots, respectively, giving the Cauchy horizon ($r_-$) and the event horizon ($r_+$) (see Figures~\ref{fig:parameterSpace} and \ref{fig:HorizonPlot}). Interestingly, both $r_-$ and $r_+$ are smaller than the the respective horizons of the Kerr BHs ($k=0$), $r_{\mp}^{Kerr} = M\mp\sqrt{M^2-a^2}$. 
It turns out, there exists an extremal spin $a_E$ for each $k$, such that $\Delta=0$ has a double root corresponding to an extremal BH with degenerate horizons. For $a<a_E$ $\Delta=0$ has two simple roots and no roots for $a>a_E$, which respectively correspond to a BH with Cauchy and event horizons, and a NS.
Interestingly, for a given $a\, (\gtrsim 0.855M)$ there exist two extremal values of PFDM parameter such that when $k=k_{E1}$ or $k=k_{E2}$, $\Delta=0$ admits a double root corresponding to an extremal BH, whereas, when $k_{E1}<k<k_{E2}$, a NS is obtained and $k<k_{E1}$ or $k>k_{E2}$ correspond to BHs with two horizons (cf. Figure~\ref{fig:HorizonPlot}).
The shaded region in the parameter space (see Figure~\ref{fig:parameterSpace}) correspond to BHs with two horizons, and the points on the red line in Figure~\ref{fig:parameterSpace} correspond to extremal rotating PFDM BHs.
Further, the extremal values of the spin $a_E$ is dependent on $k$, viz., the extremal rotating PFDM BH occurs for $k_{E1}=0.086M$, $k_{E2}=0.525M$ at $a=0.9M$ and for $a_E=0.525M$; whereas, the extremal $k_E$ is dependent on $a$, for e.g., $a_E=0.525M$ at $k=0.6M$ (cf. Figure~\ref{fig:HorizonPlot}). 

Interestingly, the metric (\ref{metric}), like Kerr metric, is independent of coordinates $t$ and $\phi$, and hence admits killing vectors $\eta_{(t)}^\mu = \delta_t^\mu$ and $\eta_{(\phi)}^\mu = \delta_{\phi}^\mu$, which are associated with the conserved quantities energy $\mathcal{E}$ and $\mathcal{L}$ respectively \citep{Carter:1968rr,Kumar:2018ple,Afrin:2021imp}; we shall see that these constants of motion significantly simplify solving the coupled null geodesic equations.
The photon motion in the spacetime (\ref{metric}) is governed by the Hamilton-Jacobi equation \citep{Carter:1968rr,Ghosh:2022kit}, 
\begin{eqnarray}
\label{HamJam}
\frac{\partial S}{\partial \tau} = -\frac{1}{2}g^{\alpha\beta}\frac{\partial S}{\partial x^\alpha}\frac{\partial S}{\partial x^\beta} ,
\end{eqnarray}
where $\tau$ is the affine parameter, and $S$ is the Jacobi action, which is given by
\begin{eqnarray}
S=-{\cal E} t +{\cal L} \phi +S_r(r)+S_\theta(\theta) \label{Jacobi},
\end{eqnarray}
where $S_r(r)$ and $S_r(\theta)$, respectively, are functions of coordinates $r$ and $\theta$ only.
We apply the approach by Carter \citep{Carter:1968rr, Chandrasekhar:1985kt} to obtain the decoupled null geodesic equations in first order differential form as \citep{Carter:1968rr, Chandrasekhar:1985kt,Kumar:2017tdw,Afrin:2021imp}
\begin{eqnarray}
\Sigma \frac{dt}{d\tau} &=& \frac{r^2 + a^2}{\Delta}[\mathcal{E}(r^2+a^2)-a\mathcal{L}]-a(a\mathcal{E}\sin^2{\theta}-\mathcal{L}),\label{geo1}\\
\Sigma \frac{dr}{d\tau} &=& \pm\sqrt{\mathcal{R}(r)},\label{geo2}\\
\Sigma \frac{d\theta}{d\tau} &=& \pm\sqrt{\Theta(\theta)},\label{geo3}\\
\Sigma \frac{d\phi}{d\tau} &=& \frac{a}{\Delta}[\mathcal{E}(r^2+a^2)-a\mathcal{L}]-\left(a\mathcal{E}-\frac{\mathcal{L}}{\sin^2{\theta}}\right).\label{geo4}
\end{eqnarray}
where $\mathcal{R}(r)$ and $\Theta(\theta)$ are related to the effective potential for radial and polar motions, and are given as
\begin{eqnarray}
\mathcal{R}(r) &=& (\mathcal{E}(r^2 + a^2)-a\mathcal{L})^2 - \Delta ((a\mathcal{E}-\mathcal{L})^2 + \mathcal{K}),\label{R(r)}\\
\Theta(\theta) &=& \mathcal{K} - \left(\frac{\mathcal{L}^2}{\sin^2{\theta}}-a^2\mathcal{E}^2\right)\cos^2{\theta}.\label{Theta(theta)}
\end{eqnarray}
The separability constant $\mathcal{K}$ is related to a hidden symmetry of the Petrov type D  metric (\ref{metric}), the Carter constant $\mathcal{Q}$ \citep{Carter:1968rr,Hioki:2008zw}, through $\mathcal{K}=\mathcal{Q}-(a\mathcal{E}-\mathcal{L})^2$ \citep{Carter:1968rr,Chandrasekhar:1985kt}, where $\mathcal{K}$ determines the latitudinal motion of the photon. 
The geodesic worldline of the rotating PFDM BHs can be completely determined by the first integrals of motion, $\mathcal{E}$, $\mathcal{L}$ and $\mathcal{K}$. The only bound photon trajectories are those for which $r =$ constant \citep{Carter:1968rr, Chandrasekhar:1985kt,Kumar:2017tdw,Afrin:2021ggx,Afrin:2022ztr} and further, all photons, regardless of their energy, follow the same trajectories \citep{Chandrasekhar:1985kt,Teo:2020sey}. One can reduce the degrees of freedom by introducing two dimensionless parameters
\begin{eqnarray}\label{new_parameter}
\eta=\mathcal{K}/\mathcal{E}^2,\;\;\;\;\; \xi=\mathcal{L}/\mathcal{E}.
\end{eqnarray}
and thus parameterise the orbits in two parameters instead of three \citep{Kumar:2018ple, Kumar:2020yem}. Due to spacetime symmetries, geodesics along $t$ and $\phi$ do not result in non-trivial orbits. Rewriting Eq. (\ref{Theta(theta)}) in terms of $u = \cos{\theta}$, we get
\begin{eqnarray}
\Theta(u) = \eta - (\eta + \xi^2 - a^2)u^2 -a^2u^4.\label{Theta_in_u}
\end{eqnarray}
Obviously, $\eta \geq 0$ is required for possible $\theta$ motion, hence to get the physically allowed region, we must have $\Theta(u) \geq 0$ \citep{Carter:1968rr, Chandrasekhar:1985kt}. Thus, boundaries of the physically allowed region can be obtained by solving $\Theta(u) = 0$ for $u^2$, and for this purpose we consider three cases viz. $\eta > 0$, $\eta = 0$ and $\eta < 0$.
Further, by setting $u = 1$, we get $\Theta(u) = -\xi^2$, which is negative, and hence is outside the allowed range of solutions.
\paragraph{Case $\eta = 0$:} 
In this case, $\Theta(u)=0$ simplifies to 
\begin{eqnarray}\label{u_eqn_simplified}
- (\xi^2 - a^2)u^2 -a^2u^4 = 0,
\end{eqnarray}
which admits two non-negative roots $u_0^2 = 0$ and $u_0^2=1-\xi^2/a^2$. When $\xi^2 < a^2$ both roots are relevant, and $u_0$ shift to higher $u$ as $\xi$ increases (illustrated in the left panel of Fig. \ref{fig:u_eqn_1}). However, when $\xi^2 > a^2$ only the first root exists (cf. right panel of Fig. \ref{fig:u_eqn_1}). Both these solutions give rise to equatorial orbits, however, the latter is stable under perturbation while the former is not.

\begin{figure}[t]
	\centering
	\includegraphics[width=0.45\textwidth]{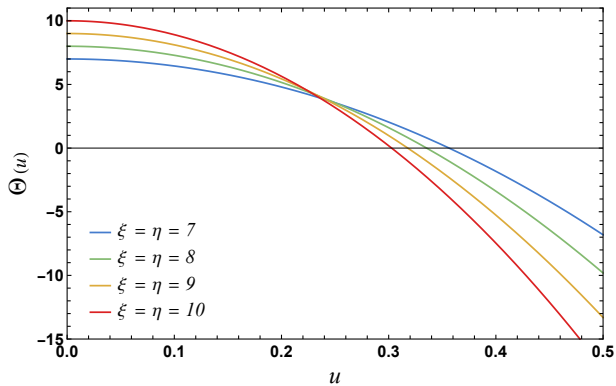}
    \caption{The variation of $\Theta(u)$ when $\eta > 0$. The intersection point shifts to a lower $u$ as $\xi$ increases.}
    \label{fig:u_eqn_2}
\end{figure}

\paragraph{Case $\eta > 0$:} $\Theta(u)=0$ immediately solves to yield
\begin{eqnarray}\label{u_eqn_solution}
u_0^2 = \frac{(a^2 - \eta - \xi^2) + \sqrt{(a^2 - \eta-\xi^2)^2+4a^2\eta}}{2a^2},
\end{eqnarray}
where $-|u_0| < u < +|u_0|$. In this case, the behaviors of $\Theta(u)$ is depicted in Fig. \ref{fig:u_eqn_2} where we plot the general shape of $\Theta(u)$. Clearly, the latitudinal turning point ($\Theta(u)=0$) is at a lower value of $u$ and hence a larger $\theta$, as $\xi$ and $\eta$ increase. 
%For a given $a$, as $k$ increases, the radius of the prograde and retrograde orbits decreases to some extent before they start to increase again. But, the radius of the prograde orbit decreases, while the radius of retrograde orbit increases with the increase in $a$ for a given $k$ (cf. Figure~\ref{fig:u} and Tables~\ref{Table:r1r2_1}, \ref{Table:r1r2_2}).

Henceforth, we shall restrict ourselves to $\eta \geq 0$, and when $\eta < 0$, we find that for the solution $u_0^2$ to be non-negative in Eq. (\ref{Theta_in_u}) $ a^2 - \eta - \xi^2 > 0.$, which is rather restrictive and is ignored in our analysis.
For the spherically symmetric BHs, all circular orbits are planar, i.e., orbits with $\dot{\theta}=0$, however, as demonstrated, the frame dragging effects lead to non-planar bound orbits in rotating spacetimes.
%%%%%%%%%%%%%%%%%%%%%%%%%%%%%%%%%%%%%%%
%\section{Spherical Photon Orbits around rotating PFDM BHs} \label{sect:SphericalOrbitsPFDM}

The spherical photon orbits, with constant radial coordinate, are not confined to latitudinal oscillations \citep{Carter:1968rr}, and have been analysed earlier for BHs \citep{Cunha:2017eoe,Charbulak:2017bpj,Teo:2020sey,Tavlayan:2020cso,Fathi:2022ntj,Das:2021otl} and NSs \citep{Liang:1974ha,Charbulak:2018wzb,Potashov:2019kxq}.
%%%%%%%%%%%%%%%%%%%%%%%%%%%%%%%%
\begin{table}
    \caption{Values of $r_+$, $r_p^-$, $r_p^*$ and $r_p^+$ }
    \begin{subtable}{.5\linewidth}
      \centering
        \caption{$a=0.9M$, $k\in[0,0.086M)\cup (0.525M, 0.8M]$}
       \begin{tabular}{|ccccc|}
\hline $k/M$ & $r_+/M$ & $r_p^-/M$ & $r_p^*/M$ & $r_p^+/M$\\ \hline
0 & 1.4343 & 1.5558 & 2.5582 & 3.9087 \\
 0.05 & 1.1233 & 1.1695 & 2.2348 & 3.6115 \\
 0.6 & 0.8289 & 0.8536 & 1.7715 & 2.8632 \\
 0.7 & 0.9302 & 0.9778 & 1.8237 & 2.8585 \\
 0.8 & 1.0115 & 1.0803 & 1.8836 & 2.8693 \\ \hline
\end{tabular}
    \end{subtable}
    \begin{subtable}{.5\linewidth}
      \centering
        \caption{$k=0.6M$, $a<a_E=0.9202M$}
        \begin{tabular}{|ccccc|}
\hline $a/M$ & $r_+/M$ & $r_p^-/M$ & $r_p^*/M$ & $r_p^+/M$\\ \hline
 0 & 1.4645 & 2.1508 & 2.1509 & 2.1510 \\
 0.3 & 1.4198 & 1.8529 & 2.1164 & 2.4111 \\
 0.5 & 1.3333 & 1.6196 & 2.0518 & 2.5702 \\
 0.7 & 1.17908 & 1.3328 & 1.9447 & 2.7204 \\
 0.9 & 0.8289 & 0.8537 & 1.7715 & 2.8633 \\ \hline
\end{tabular}
    \end{subtable} 
\label{Table:r1r2_1}
\end{table}
%%%%%%%%%%%%%%%%%%%%%%%%%%%%%%%%%%%%%%%%%%%%%%%%%%%%%%%%%%%%
To see the effects of surrounding PFDM we rewrite Eq.~(\ref{R(r)}) in terms of $\eta$ and $\xi$ as
\begin{eqnarray}\label{Rr_eta_xi}
\mathcal{R}(r) = (r^2 + a^2 -a\xi)^2 - \Delta [\eta + (\xi -a )^2].
\end{eqnarray}
The unstable spherical photon orbits with radii $r=r_p$ satisfy \citep{Chandrasekhar:1985kt}
\begin{equation}
\left.\mathcal{R}\right|_{(r=r_p)}=\left.\frac{\partial \mathcal{R}}{\partial r}\right|_{(r=r_p)}=0, \quad \left.\frac{\partial^2 \mathcal{R}}{\partial r^2}\right|_{(r=r_p)}\leq 0, \label{tobesolved} 
\end{equation}
solving which yields the critical impact parameters
\begin{align}
\eta_{c}=&\frac{r^2 \left(8 \Delta (r) \left(2 a^2+r \Delta '(r)\right)-r^2 \Delta '(r)^2-16 \Delta (r)^2\right)}{a^2 \Delta '(r)^2},\label{sol2a}\\
\xi_{c}=&\frac{\left(a^2+r^2\right) \Delta '(r)-4 r \Delta (r)}{a \Delta '(r)}.\label{sol2b}
\end{align}
The Eqs. (\ref{sol2a}) and (\ref{sol2b}) get modified due to surrounding PFDM; in its absence ($k$=0), $m(r)=M$ and the above equations reduce to that of the Kerr BHs \citep{Teo:2020sey, Tavlayan:2020cso}
\begin{eqnarray}
\eta_c^{k} &=& -\frac{r^3(r^3-6Mr^2+9M^2r-4a^2M)}{a^2(r-M)^2}, \label{sol2a_kerr}\\
\xi_c^{k} &=& -\frac{r^3 -3Mr^2+a^2r+a^2M}{a(r-M)}.\label{sol2b_kerr}
\end{eqnarray}
%%%%%%%%%%%%%%%%%%%%%%%%%%%%%%%%%%%%%%%
\begin{figure*}[t]
\centering
\includegraphics[width=0.47\textwidth]{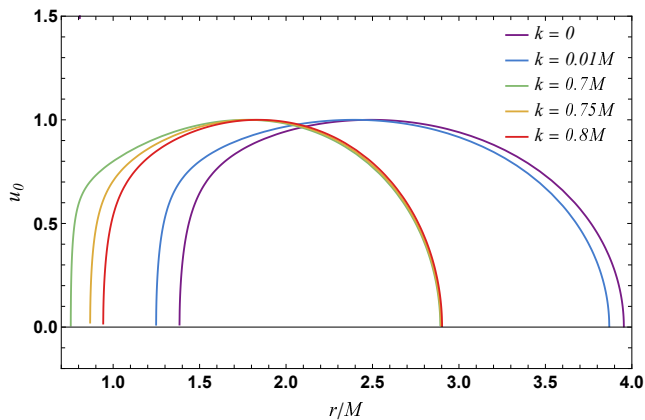} \hspace{0.3cm}
\includegraphics[width=0.47\textwidth]{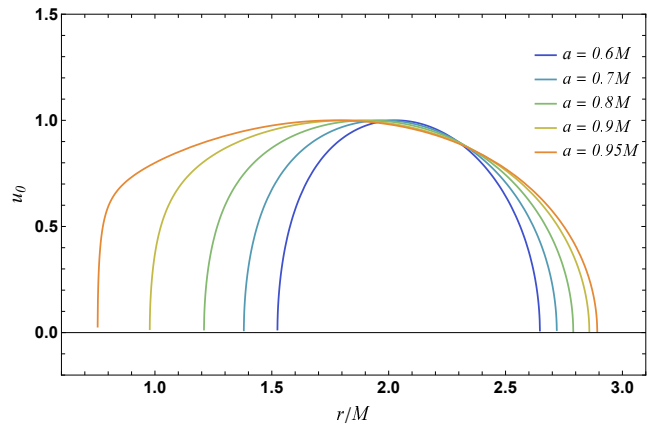}
\caption{Plots for $u_0$ vs  $r$ using Eq.~\ref{u_eqn_solution}. The points of intersection of the curve on the horizontal axis give the radii of the prograde and retrograde orbit (See also Table \ref{Table:r1r2_1}).} \label{fig:u}
\end{figure*}
%%%%%%%%%%%%%%%%%%%%%%%%%%%%%%%%%%%%%
Photons travelling in unstable spherical orbits can either co-rotate with the BH (prograde) or counter-rotate (retrograde), and their radii $r_p^{\mp}$ can be, respectively, identified as the real positive solutions of $\eta_c = 0$. For the Kerr BH, photon orbit radii are an explicit function of BH spin and fall into the range $M \leq r_p^- \leq 3M$ and $3M \leq r_p^+ \leq 4M$, due to the Lense-Thirring effect. For $\eta_c >0$ the photon region is a three-dimensional spherical shell, $\mathcal{A}$: $r\in[r_p^-, r_p^+]$, $\theta\in[\cos^{-1}({-|u_0|}), \cos^{-1}({|u_0|})]$, $\phi\in[0, \pi]$, where $u_0$ is given by Eq.~(\ref{u_eqn_solution}) (see Figure~\ref{fig:PhotonRegion}). Although spinning BHs have two separate photon zones, one inside the Cauchy horizon ($r_p<r_-$) and the other outside the event horizon ($r_p>r_+$), we will be focusing solely on the latter for our analysis. Further, $\xi_c$ is a monotonically decreasing function of $r_p$, with $\xi_c(r_p^-)>0$ and $\xi_c(r_p^+)<0$, whereas $\xi_c$ is vanishing at $r_p = r_p^*$ where the prograde orbits change to retrograde ones $(r_p^-<r_p^*<r_p^+)$. Solving Eq.~(\ref{sol2a}) for the given values of $k$ and $a$, we obtain the radii of photon orbits (cf. Tables~\ref{Table:r1r2_1}).

\begin{figure}[t]
\centering
\includegraphics[width=0.45\textwidth]{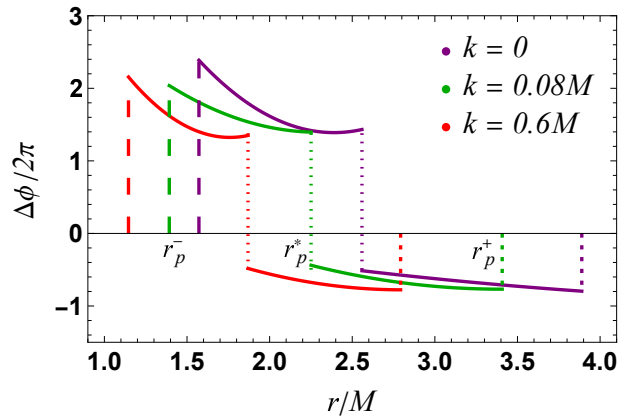}
\caption{Plot for $\Delta \phi$/2$\pi$ vs  $r$ for  $k = 0$, $k=0.08M$, and $k = 0.6M$. $\Delta \phi$/2$\pi>0$ represents the prograde orbits while $\Delta \phi$/2$\pi<0$ shows the retrograde orbits. The radii $r_p^-$ and $r_p^+$ are shown numerically in Table \ref{Table:r1r2_1}.}   \label{fig:delphi}
\end{figure}

To further analyse the spherical photon orbits around rotating PFDM spacetimes, we note that they have latitudinal oscillation and it is useful to measure the resulting azimuthal periodicity. For this, we consider the change in azimuthal angle $\Delta \phi$ for one complete latitudinal oscillation, which is given by
\begin{eqnarray}
\Delta \phi =\oint d \phi = \int \frac{d \phi}{d \theta}d \theta = \int \frac{d \phi / d \tau}{d \theta / d \tau}d\theta. \label{Delphi_eqn}
\end{eqnarray}
For rotating PFDM BHs, on inserting Eqs. (\ref{geo3}) and (\ref{geo4}) in Eq. (\ref{Delphi_eqn}), we obtain
\begin{eqnarray}\label{Delphi_eqn_integration}
\Delta \phi(r) = \int_{0}^{2\pi} \left( \frac{\frac{a \left(a^2+r^2-a \xi \right)}{\Delta(r)}-\left(a-\frac{\xi }{\sin ^2\theta }\right)}{\sqrt{\eta -\cos ^2\theta \left(\frac{\xi ^2}{\sin ^2 \theta}-a^2\right)}} \right)d\theta.
\end{eqnarray}
For the prograde orbits, $\Delta \phi > 2\pi$, while for the retrograde orbits, $|\Delta \phi| < 2 \pi$. After numerical integration of Eq.~(\ref{Delphi_eqn_integration}) we depict the behavior of $\Delta \phi(r)$ in Figure~\ref{fig:delphi}. We note that $r_p^*$ solves the zero angular momentum equation $\xi = 0$ and $\Delta \phi$ has discontinuity at $r = r_p^*$ (cf. Figure~\ref{fig:delphi}). The intersection points on the horizontal axis $r_p^{+}$ and  $r_p^{-}$  correspond, respectively, to retrograde and prograde radius. For $r<r_p^*$, orbits are prograde with $\Delta \phi /2\pi > 0$ and for $r>r_p^*$, the orbits are retrograde with $\Delta \phi / 2 \pi < 0$. This means that the photons take more than one revolution in the $\phi$ direction to complete each latitudinal oscillation in a prograde orbit, whereas in the retrograde orbits, one latitudinal oscillation takes less than one revolution in the $\phi$ direction. Thus,  $\Delta \phi$ is increasing in prograde orbits and decreasing in the retrograde orbits. Despite the fact that photon's angular momentum is zero for orbits at $r_p^*$, they still cross the equatorial plane with a non-zero azimuthal velocity $\dot{\phi} \neq 0$ \citep{Chandrasekhar:1985kt, Wilkins:1972rs}. The influence of $k$ on $r_p^{\pm}$ and $r_p^{*}$ is clearly visible from Figures~\ref{fig:u} and \ref{fig:delphi} (see also Table~\ref{Table:r1r2_1})).

We consider the orbits with fixed $\Delta\phi$, but the photon may not be confined to one azimuthal direction. It is evident from Eq. (\ref{geo4}), $\dot{\phi}$ changes sign whenever $u^2$ takes the value 
\begin{eqnarray}\label{u1_solution}
u_1^2 &=& -\frac{r \left[k (\xi -a) \ln{ \left(\frac{r}{| k| }\right)} + 2 a M+\xi  (r-2 M)\right]}{a \left[a \xi +k r \ln{ \left(\frac{r}{| k| }\right)}-2 M r\right]}
\end{eqnarray}
Equation (\ref{u1_solution}), when $k=0$, reduces to
\begin{eqnarray}\label{u1_solution_kerr}
u_1^2 &=& \frac{r^2 (3M -r)}{a^2 (r+M)},
\end{eqnarray}
which is exactly obtained for the Kerr spacetime \citep{Chandrasekhar:1985kt,Teo:2020sey}.
\section{Visualisation of Spherical Photon Orbits} \label{sect:Visualization}
\begin{figure*}
\centering
\begin{tabular}{c c c}
     \includegraphics[width = 0.3\textwidth]{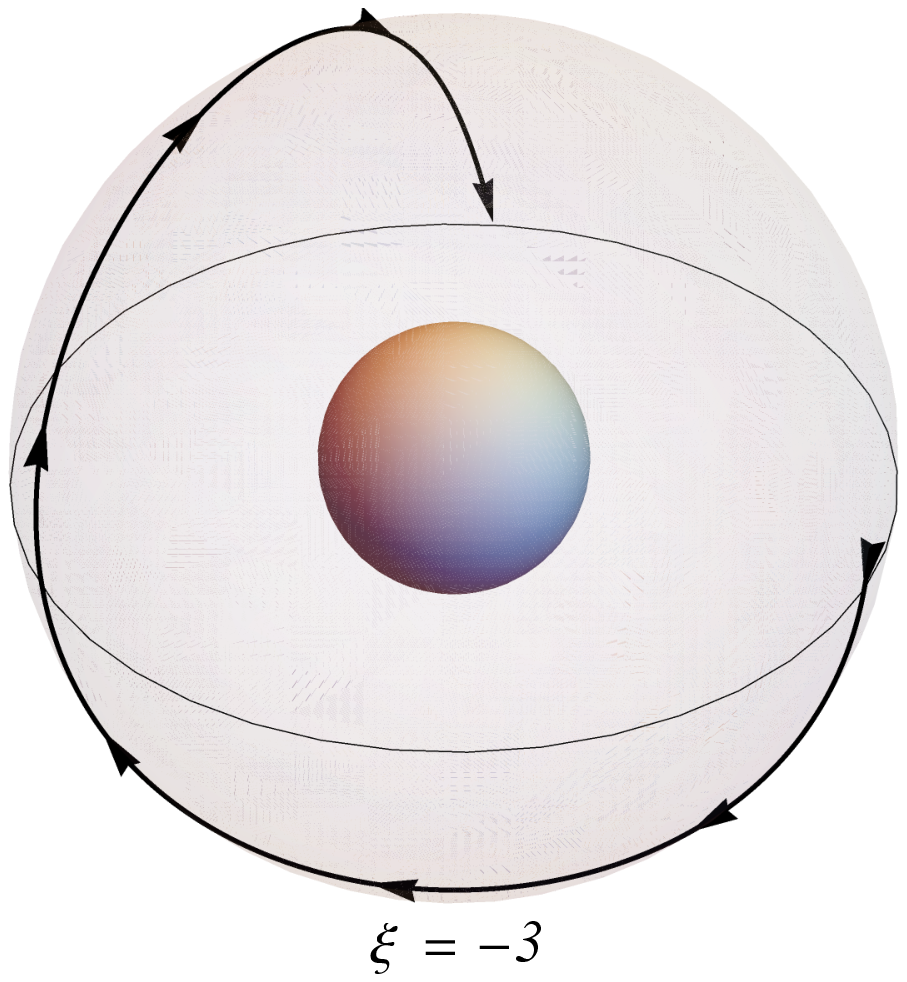}&
     \includegraphics[width = 0.3\textwidth]{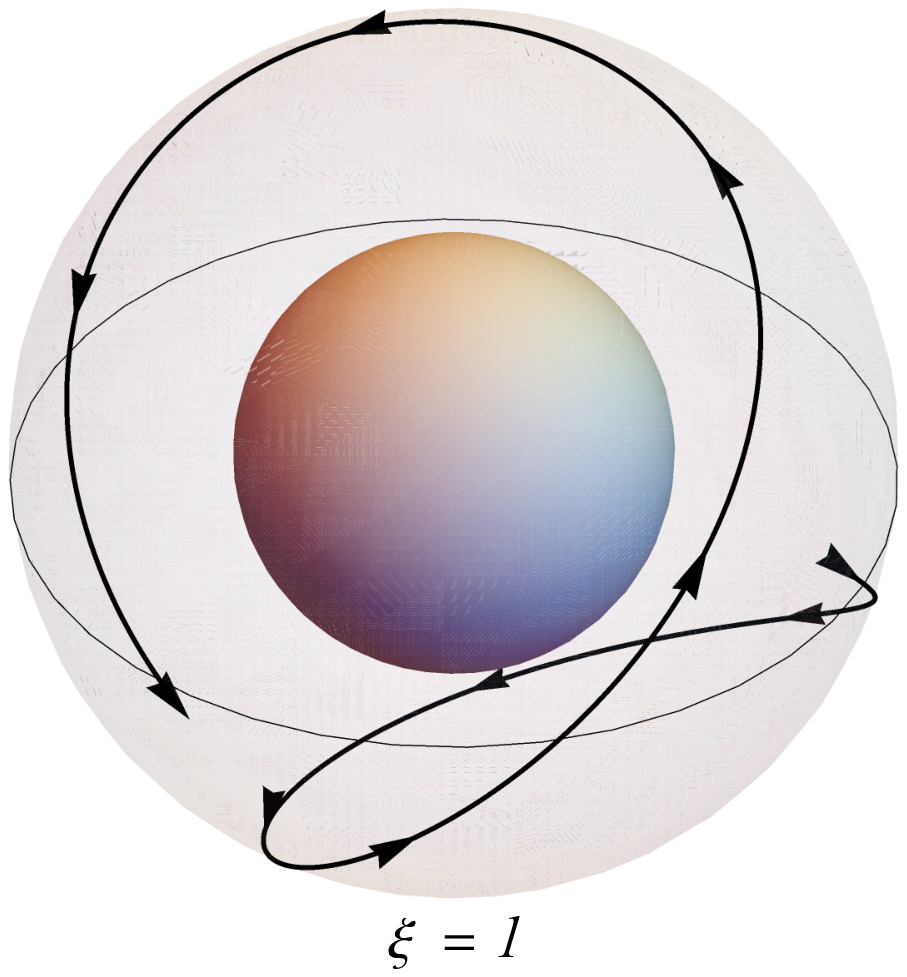}&
     \includegraphics[width = 0.3\textwidth]{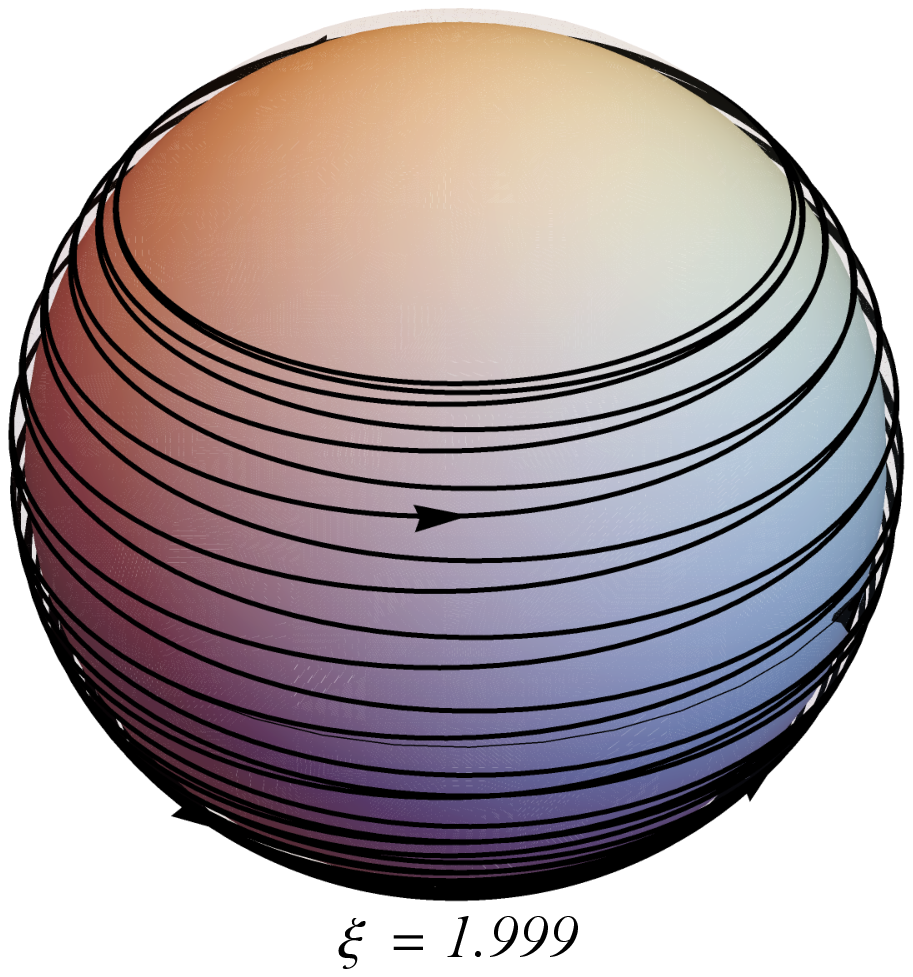}
\end{tabular}
\caption{One latitudinal oscillation of spherical photons around Kerr BH with $a\approx M$. Evidently, there is an increase in the azimuthal oscillations with $\xi$.} \label{fig:SPO1}
\end{figure*}

\begin{figure*}
\centering
\begin{tabular}{c c c}
     \includegraphics[width = 0.3\textwidth]{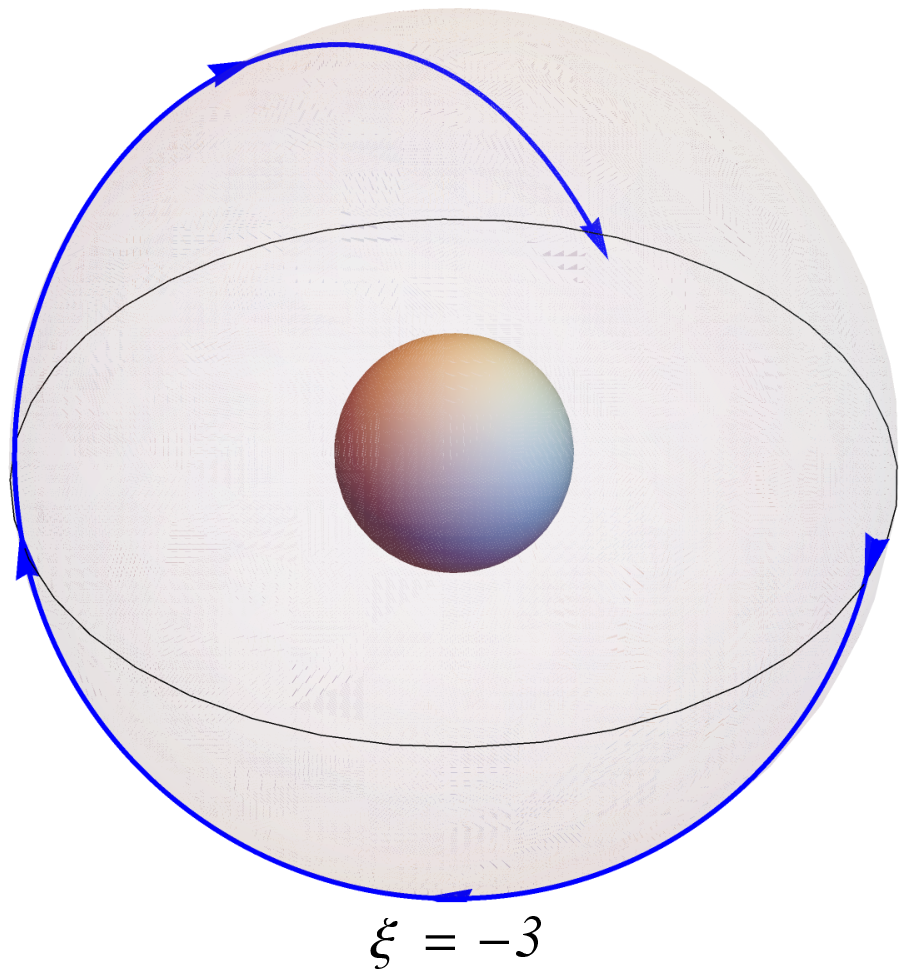}&
     \includegraphics[width = 0.3\textwidth]{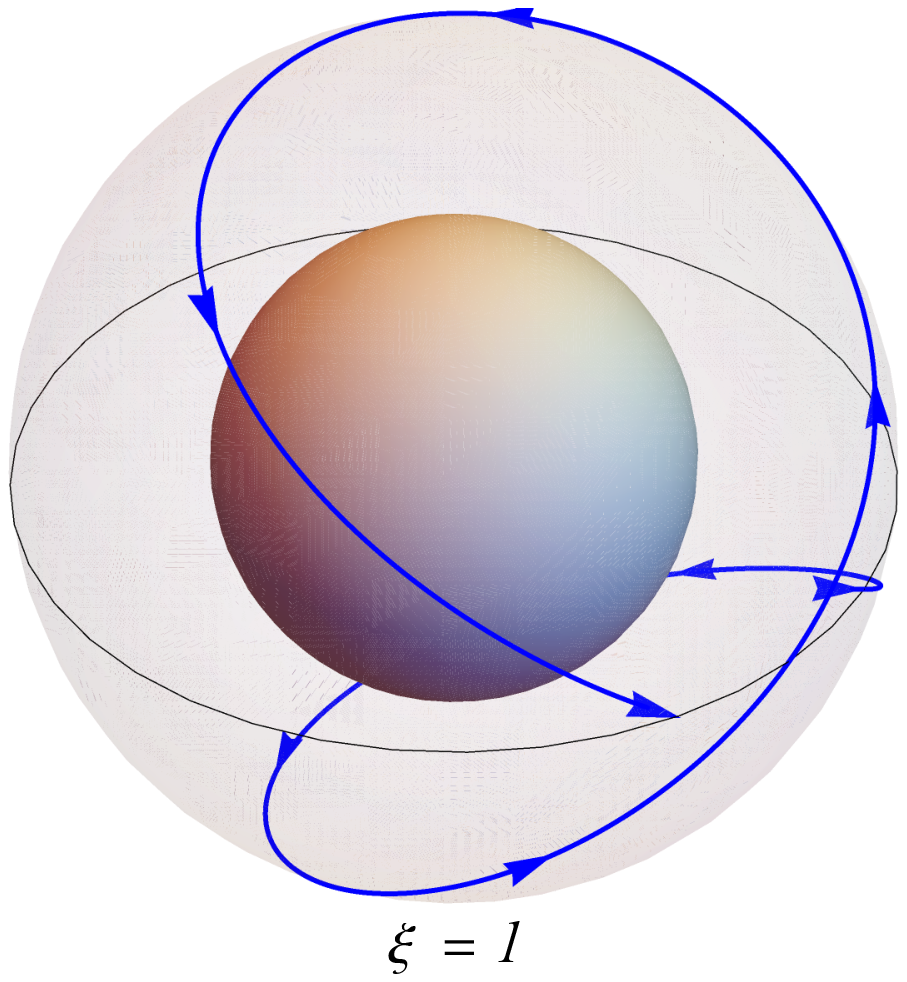}&
     \includegraphics[width = 0.3\textwidth]{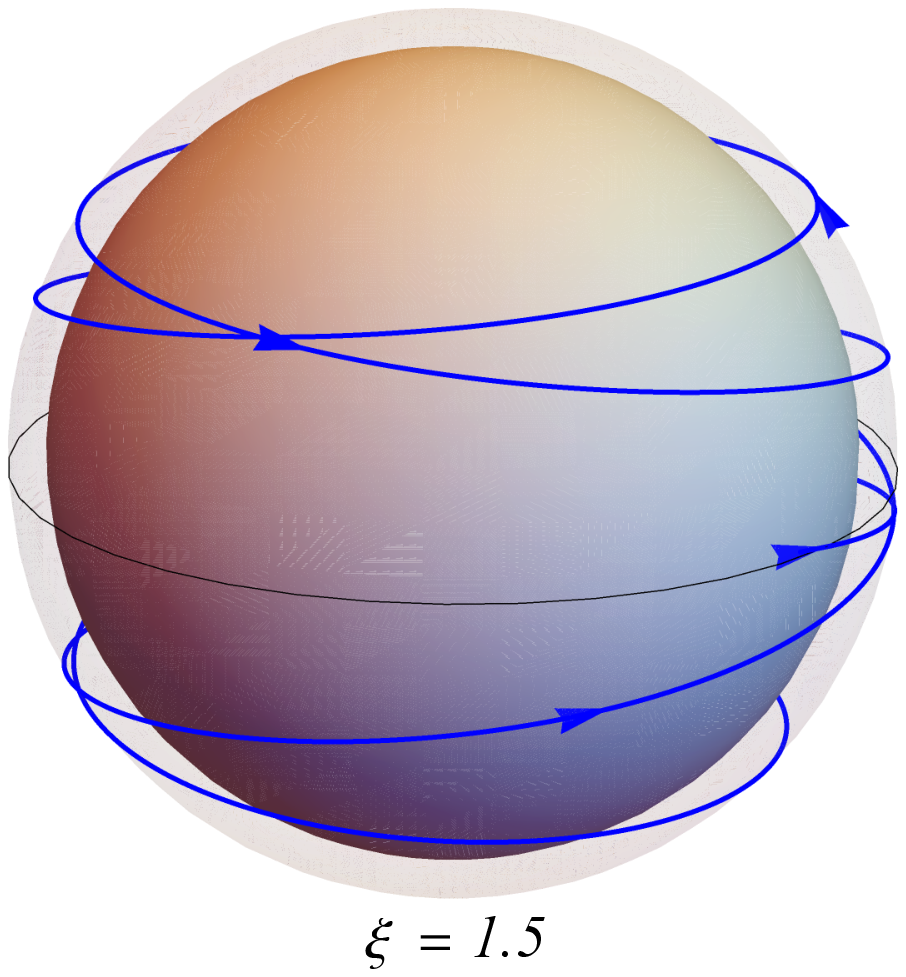}
\end{tabular}
\caption{One latitudinal oscillation of spherical photon orbits around rotating PFDM BH ($k= 0.4M$). There is no monotonic behaviour of azimuthal oscillations with increase in $\xi$.} \label{fig:SPO2}
\end{figure*}

\begin{figure*}[t]
\centering
\begin{tabular}{c c c}
     \includegraphics[width = 0.3\textwidth]{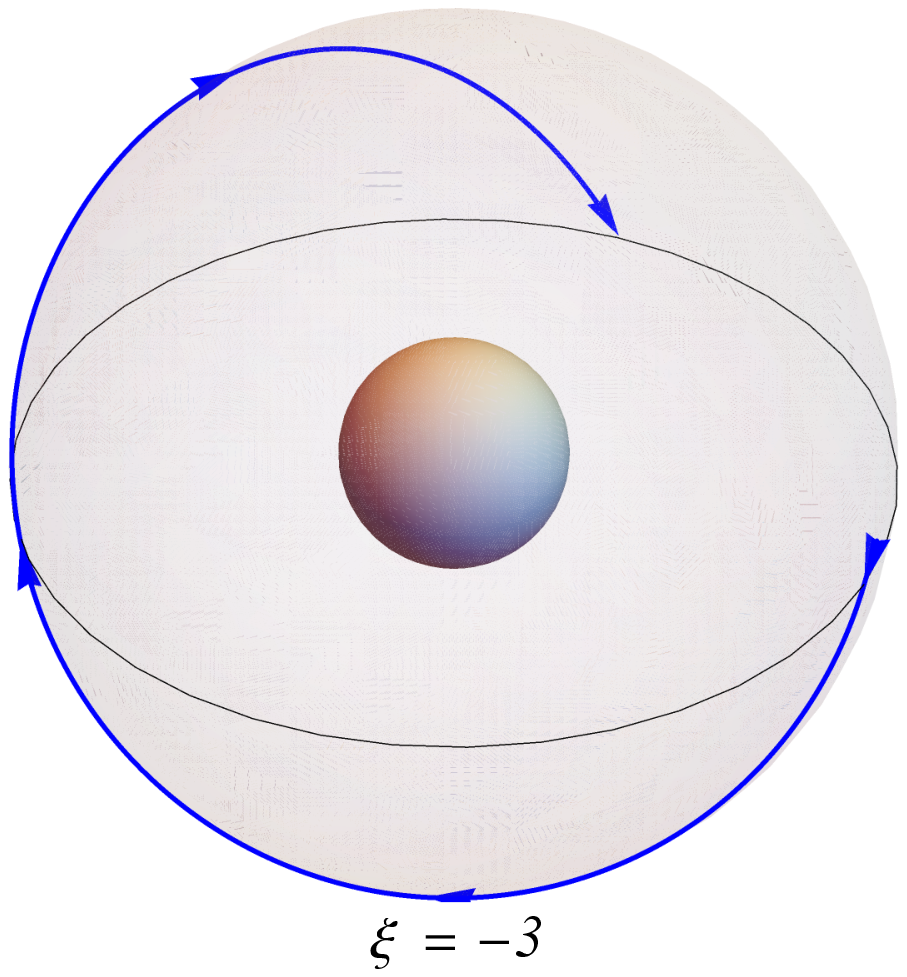}&
     \includegraphics[width = 0.3\textwidth]{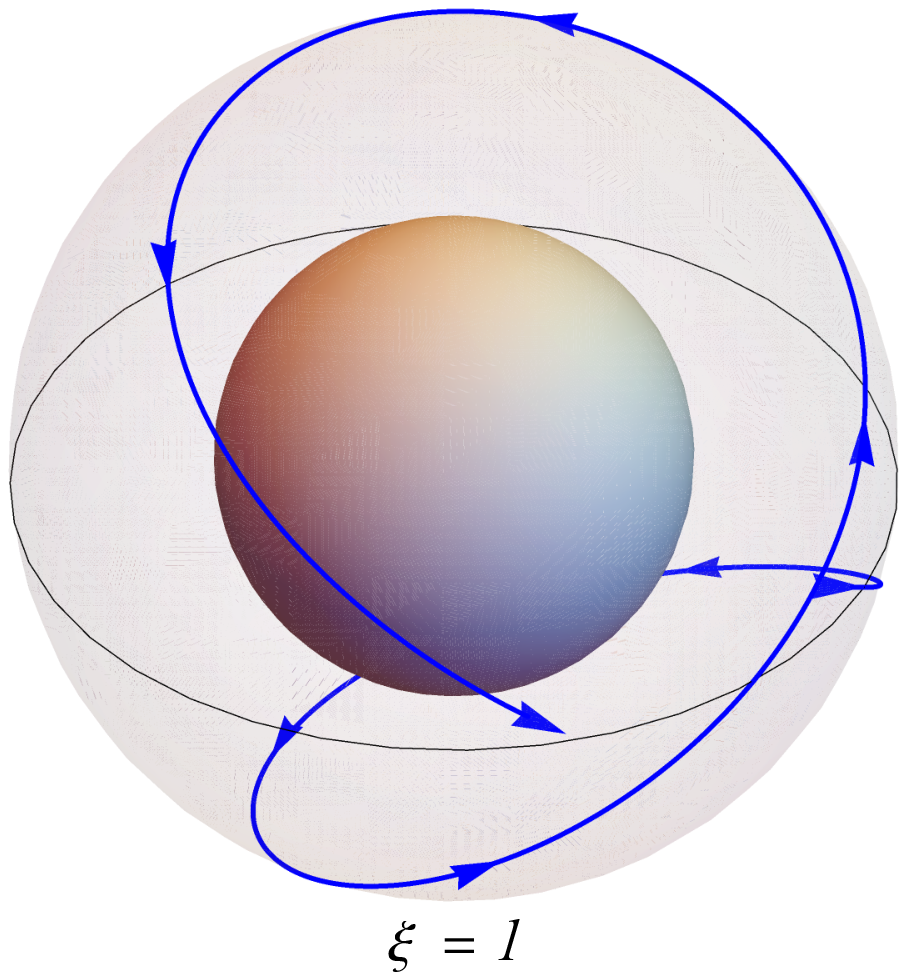}&
     \includegraphics[width = 0.3\textwidth]{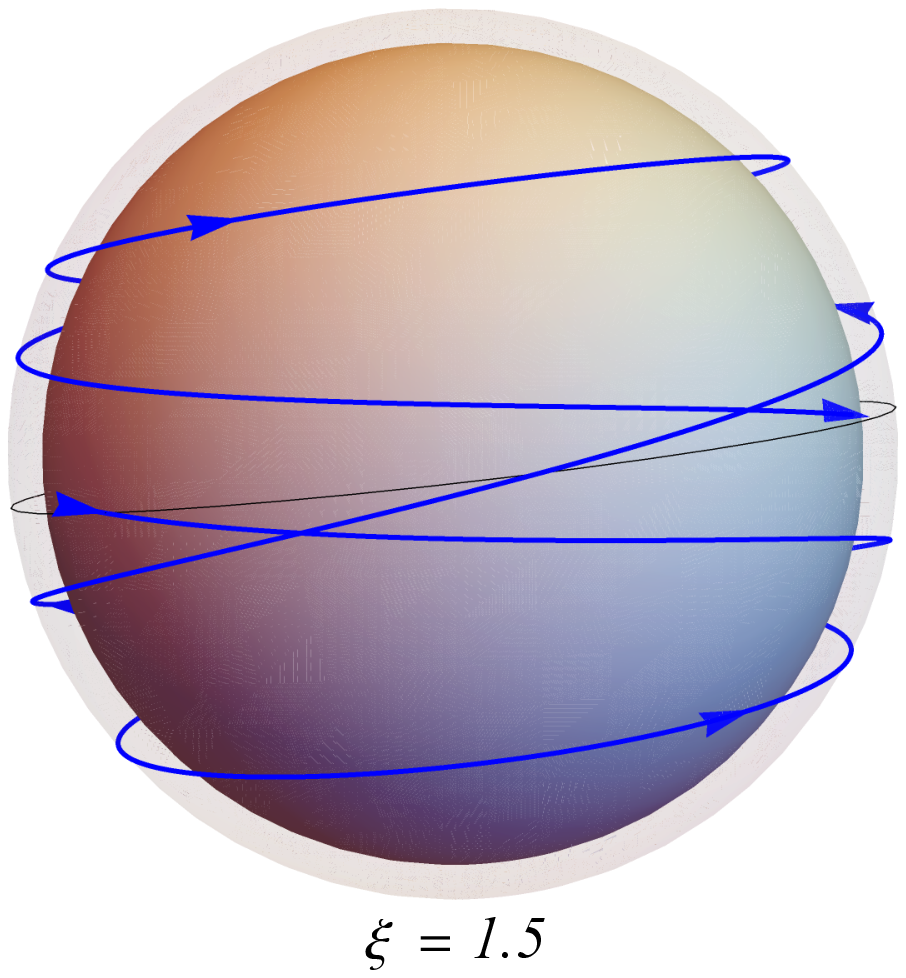}
\end{tabular}
\caption{One latitudinal oscillation of spherical photon orbits around rotating PFDM BH with $k = 0.6M$. There is no monotonic behaviour of azimuthal oscillations with increase in $\xi$.} \label{fig:SPO3}
\end{figure*}

\begin{figure*}[t]
\centering
\includegraphics[width = 0.4\textwidth]{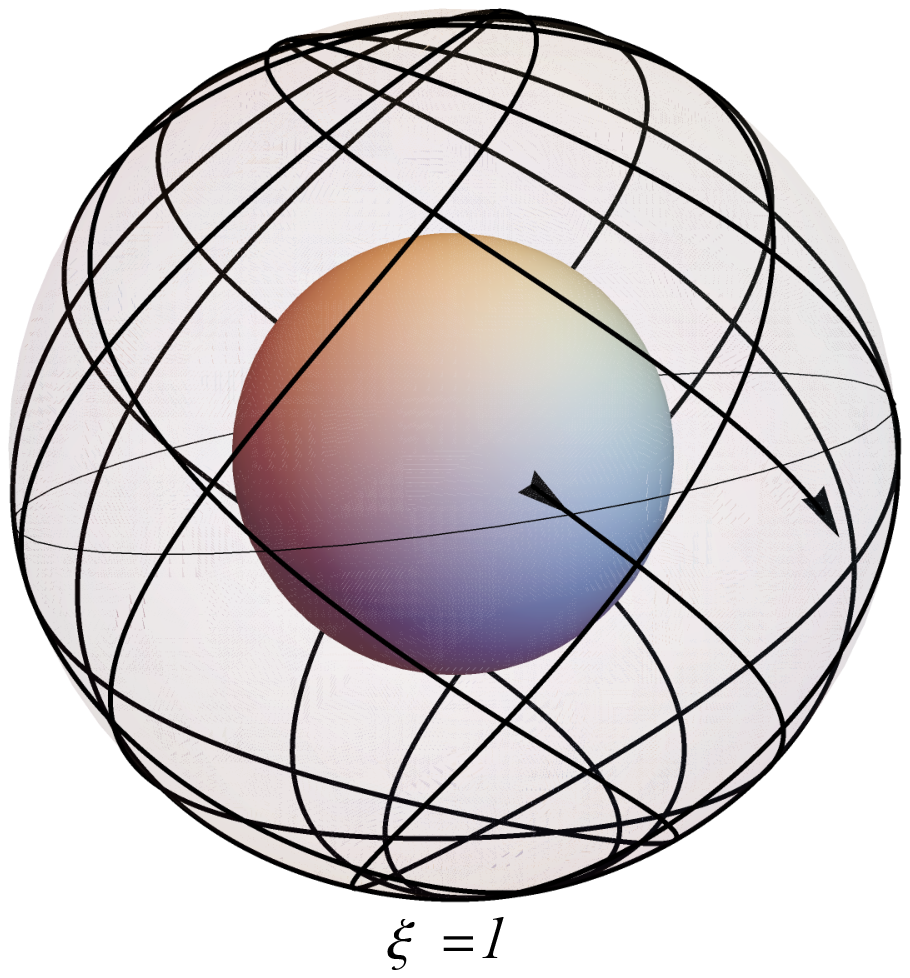} \hspace{0.5cm}
\includegraphics[width = 0.4\textwidth]{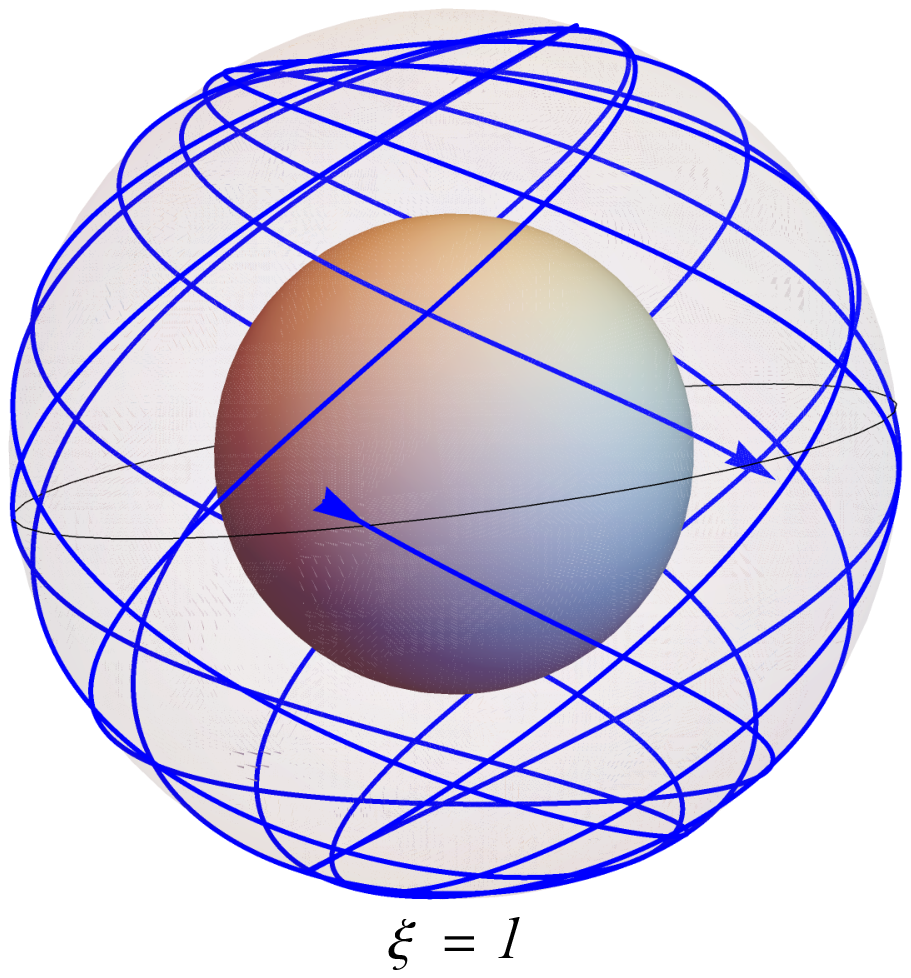} 
\caption{Seven latitudinal oscillations of spherical photon orbits around Kerr BH (left) rotating PFDM BH ($k=0.6M$) (right).} \label{fig:Pleasing}
\end{figure*}

We first compute the Hamilton-Jacobi equation (\ref{HamJam}) in terms of $M$, $\Delta$ and $\Sigma$ with the aid of Christoffel symbols for the rotating PFDM spacetime (\ref{metric}). Using the 4th order Runge-Kutta method, we solve the second order equations of motion (\ref{HamJam}) by setting numerical values of $a$ and $k$, which can be obtained from the parameter space for BH or NS (cf. Figure~\ref{fig:parameterSpace}); we consider here, only three value of PFDM parameter $k$. We also fix $\eta$ and $\xi$ for the spherical photon orbits as discussed in Section \ref{sect:EquationsofMotion}. The initial conditions can be obtained from Eqs.~(\ref{geo1})-(\ref{geo4}) by setting $t=0$, $\theta=\pi/2$ and $\phi=0$. The integration leads to numerical values of ($t$, $r$, $\theta$, $\phi$) which we plot in the Cartesian coordinates in Figures~\ref{fig:SPO1}-\ref{fig:boomerang} BHs and NSs, and the special case of photon \emph{boomerang}, i.e., we explicitly display the effect of PFDM on the spherical photon orbits. When no surrounding PFDM is present ($k=0$), the orbits around Kerr spacetimes are obtained.

Each orbit begins at the equator, and heads southwards; the BH is assumed to be rotating counter-clockwise. We also show the horizons (solid sphere) (cf. Figures~\ref{fig:SPO1}-\ref{fig:Pleasing}), which maybe useful to check how far from the horizons, the orbits get formed. 
%%%%%%%%%%%%%%%%%%%%%%%%%%%%%%%%%%%%
%%%%%%%%%%%%%%%%%%%%%%%%%%%%%%%%%%%%%%%%%
\begin{table}
\caption{\label{tab:table1}
Values of various parameters of the photon's trajectory at different $k$ around rotating PFDM BHs.}
\begin{ruledtabular}
\begin{tabular}{cccccc}
\textbf{$k/M$} & \textbf{$\xi$} & \textbf{$\eta$} & \textbf{$r/M$} & \textbf{$a/M$} & \textbf{$\Delta \phi$} \\ \hline
\multirow{6}{*}{0} & -3 & 25.8885 & 3.23607 & 1 & -5.27656 \\
 & -2 & 27 & 3 & 1 & -3.77684 \\
 & -1 & 25.8564 & 2.73205 & 1 & -2.04872 \\
 & 0 & 22.3137 & 2.41421 & 1 & 3.17612 \\
 & 1 & 16 & 2 & 1 & 10.3906 \\
 & 1.999 & 3.25898 & 1.03162 & 1 & 161.198 \\ \hline
\multirow{6}{*}{0.4} & -3 & 9.06473 & 2.58256 & 0.8679 & -3.22136 \\
 & -2 & 11.5839 & 2.35386 & 0.8679 & -4.90844 \\
 & -1 & 11.9484 & 2.08875 & 0.8679 & -5.41668 \\
 & 0 & 10.0401 & 1.76062 & 0.8679 & 3.04419 \\
 & 1 & 5.41716 & 1.2714 & 0.8679 & 9.12222 \\
 & 1.5 & 1.17515 & 0.86858 & 0.86 & 23.4401 \\ \hline
\multirow{6}{*}{0.6} & -3 & 7.52705 & 2.58465 & 0.9201 & -3.46309 \\
 & -2 & 10.224 & 2.35203 & 0.9201 & -3.46016 \\
 & -1 & 10.79 & 2.08231 & 0.9201 & -3.65801 \\
 & 0 & 9.12109 & 1.74823 & 0.9201 & 2.9089 \\
 & 1 & 4.81791 & 1.24788 & 0.9201 & 9.57698 \\
 & 1.5 & 0.87126 & 0.84673 & 0.91 & 24.358 
\end{tabular}
\end{ruledtabular}
\end{table}
For the Kerr BHs ($k=0$), the zero-angular momentum photons with ($\xi=0$, $r=r_p^*$) can rotate around the BH at constant radius due to the frame-dragging effect. We also investigate how the photon orbit changes behaviour when $\xi\neq0$. For $\xi>0$, the prograde orbits would swing around the pole, without passing through it, resulting in an increment in $\Delta \phi$ by $2\pi$; there is a monotonic increment in $\Delta\phi$ with increasing $\xi$. The retrograde orbits ($\xi<0$) on the other hand, would go near the pole and miss it; effectively there would be a decrement of $2\pi$ from $\Delta\phi$. There is again a monotonic increase in $\Delta\phi$ with $\xi$  (cf. Figure~\ref{fig:SPO1} and Table~\ref{tab:table1}).

%%%%%%%%%%%%%%%%%%%%%%%%%%%%%%%%%%%%%%%%%%
%%%%%%%%%%%%%%%%%%%%%%%%%%%%%%%%%%%%
As the surrounding PFDM parameter $k$ increases, the radii of spherical orbits as well as the azimuthal oscillations get impacted. For the rotating PFDM BHs ($k=0.6M$), the zero angular momentum photons ($\xi=0$) undergo lesser azimuthal oscillations (cf. Table~\ref{tab:table1}) as compared to the Kerr case. For the retrograde orbits, $\Delta\phi$ first increases and then decreases with $\xi$ contrary to the Kerr case (cf. Figure~\ref{fig:SPO3} and Table~\ref{tab:table1}); the retrograde orbits pass very near the pole but miss it. But the prograde orbits swing around the pole and the azimuthal oscillations monotonically increases with $\xi$. Further, for the prograde orbits the azimuthal oscillations decrease than the corresponding Kerr oscillations with increasing $k$. Similar behaviour is exhibited for $k=0.4M$ (cf. Figure~\ref{fig:SPO2}) but the $\Delta\phi$ differ (Table~\ref{tab:table1}).

The effect of PFDM on the spherical photon trajectories is found to be significant -- the radii of the orbits change as a result of the rise in $k$; the radii of both the prograde and retrograde orbits first decrease and then increase with $k$. Also, as $k$ increases, $\Delta \phi$ first increases and then decreases in the case of prograde orbits whereas for the retrograde orbits, $\Delta\phi$ first decreases and then increases with $k$ (cf. Table~\ref{tab:table1}). In stark contrast to the Kerr case ($k=0$), where the azimuthal oscillations monotonically increase with increasing $\xi$ (cf. Figure~\ref{fig:SPO1} and Table~\ref{tab:table1}), the $\Delta\phi$ shows no monotonic behaviour with $\xi$ in case of rotating PFDM BHs. The Figure~\ref{fig:Pleasing} shows how the spherical photon orbits of $\xi = 1$ would look like, if the photon continues to orbit the BH for 7 oscillations around the Kerr and rotating PFDM BHs respectively. 
\subsection{Photon orbits around NS } %\label{NakedSingularity}
\begin{figure*}
\centering
\includegraphics[width = 0.3\textwidth]{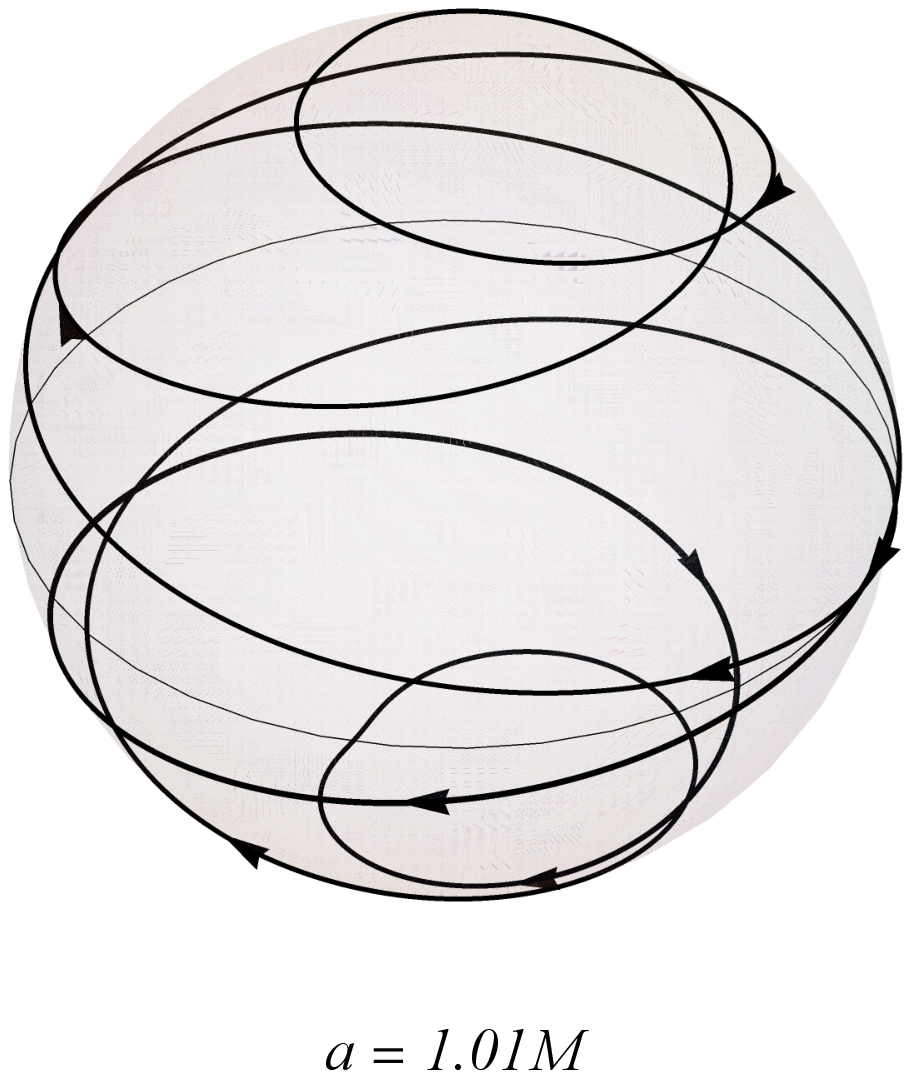} \hspace{0.25cm}
\includegraphics[width = 0.3\textwidth]{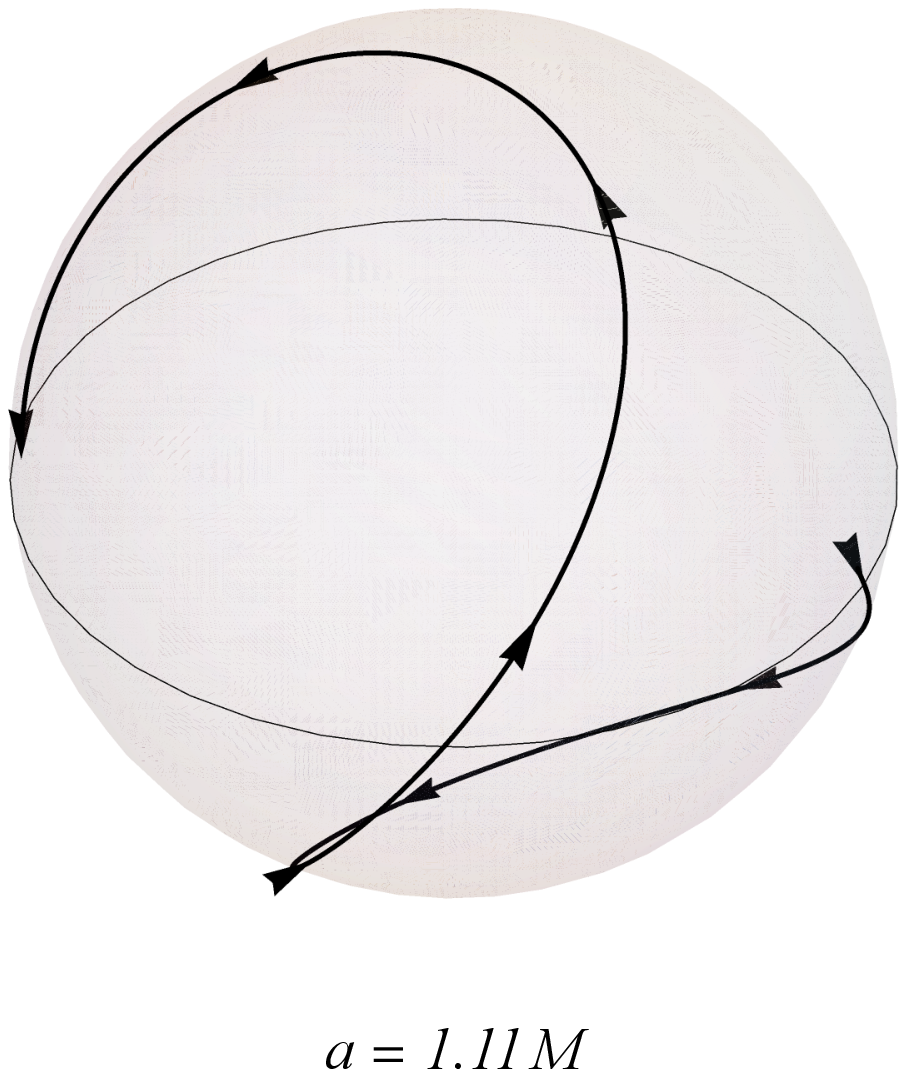} \hspace{0.25cm}
\includegraphics[width = 0.3\textwidth]{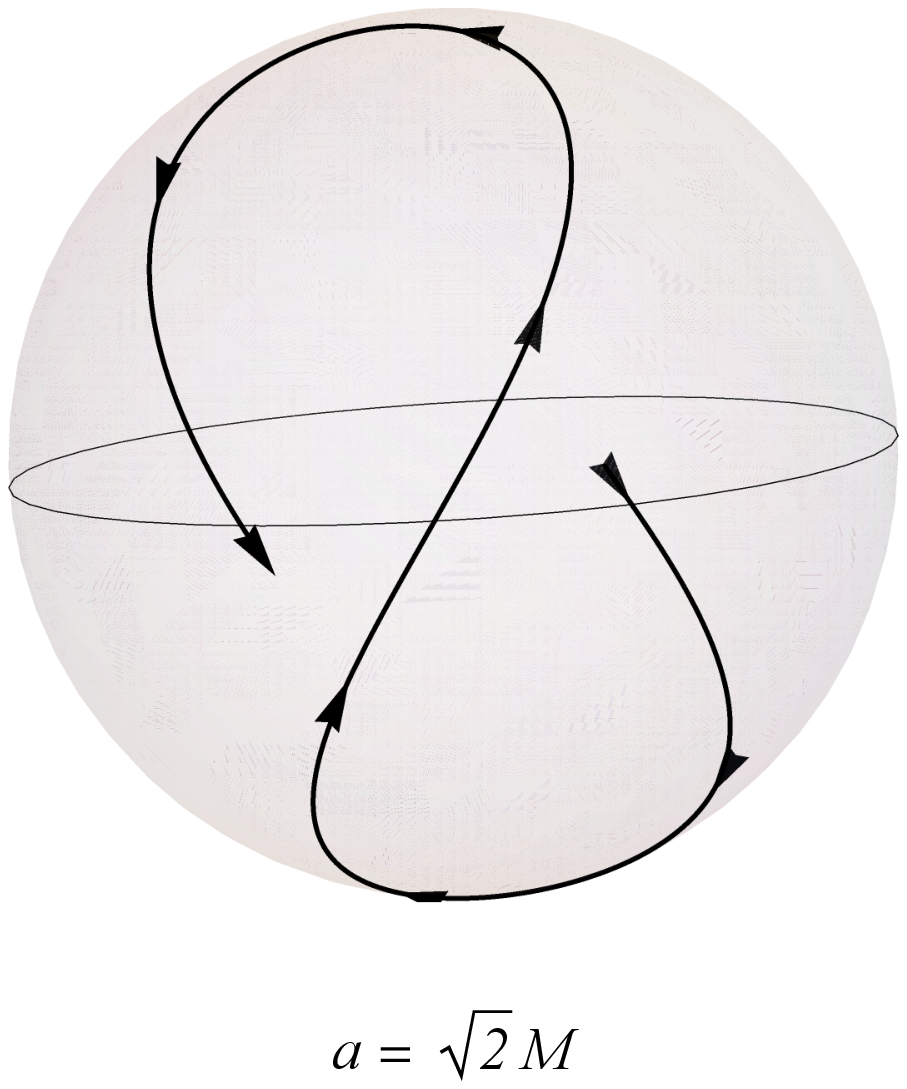}\\ 
\vspace{0.25cm}
\includegraphics[width = 0.3\textwidth]{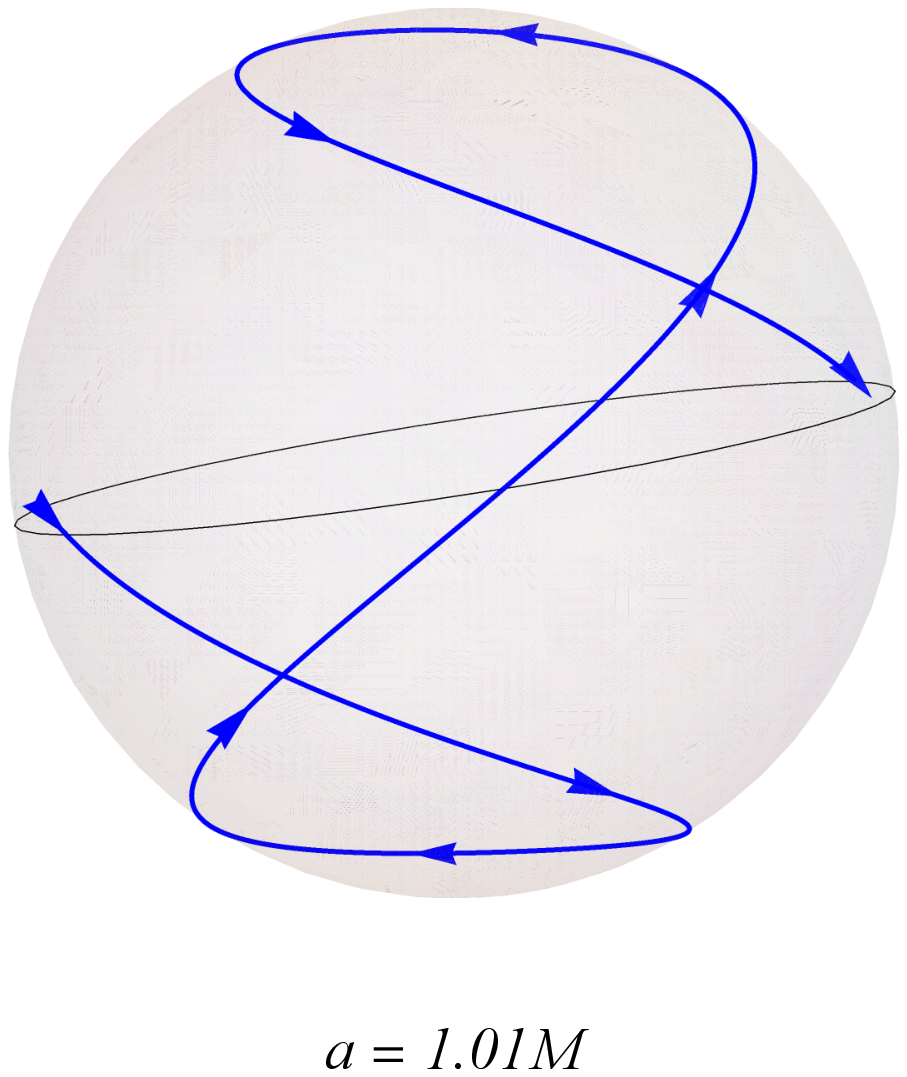} \hspace{0.25cm}
\includegraphics[width = 0.3\textwidth]{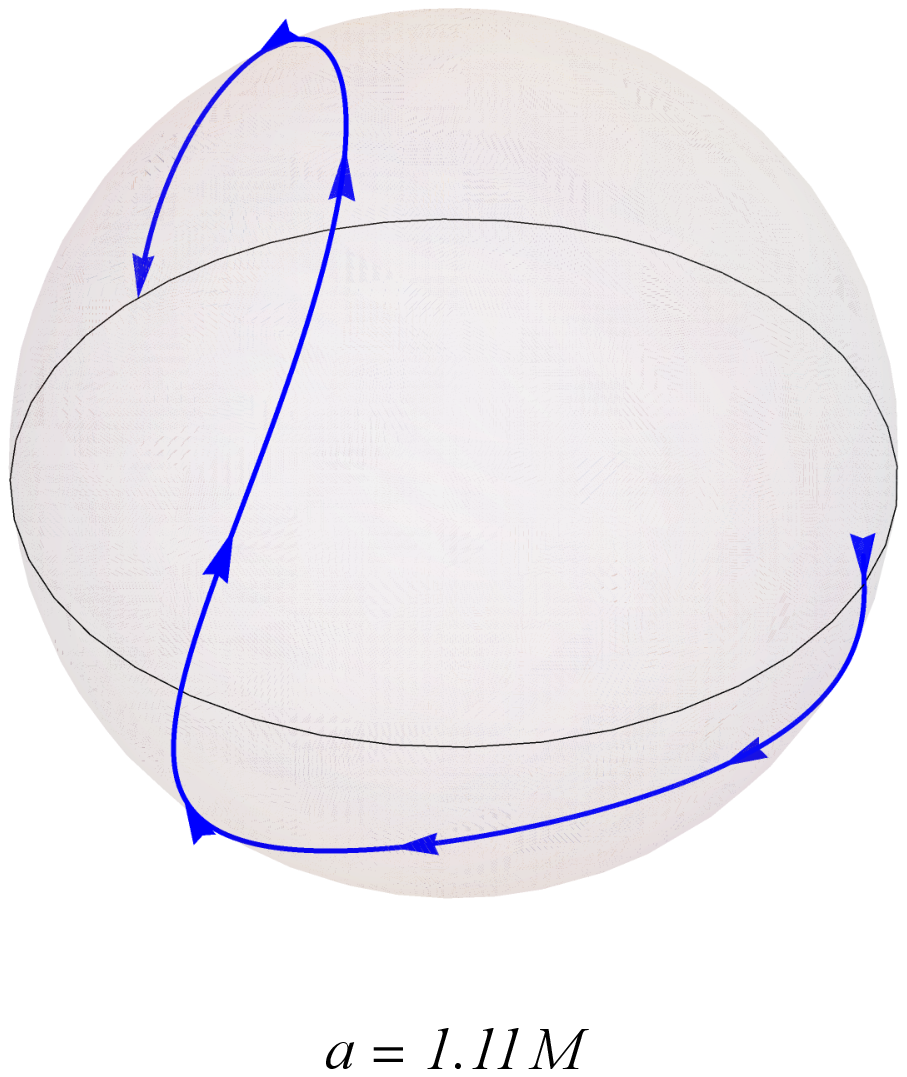} \hspace{0.25cm}
\includegraphics[width = 0.3\textwidth]{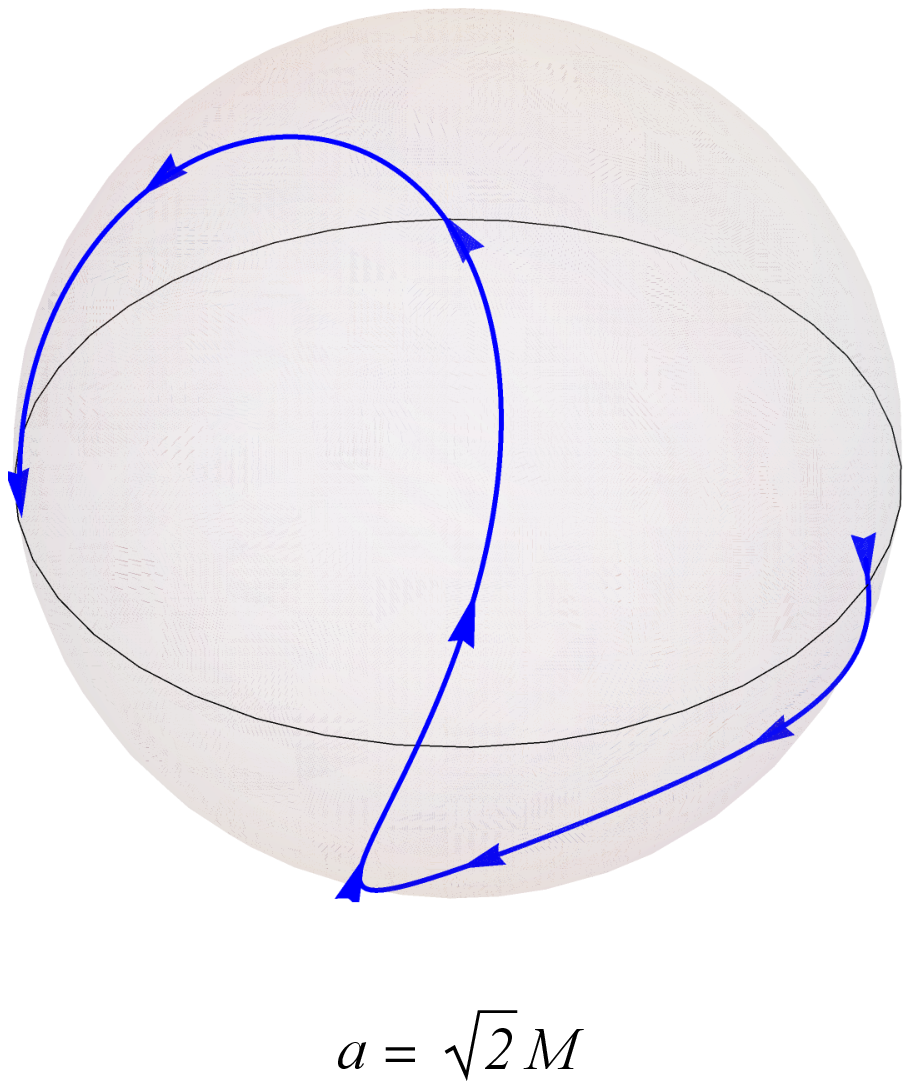}
\caption{One latitudinal oscillation in the spherical photon orbits around rotating PFDM NSs ($a>M$) for (a) $k = 0$ (top) and $k = 0.6M$ (bottom). The azimuthal oscillations decrease with $k$, see also Table.~\ref{tab:table2}.}\label{or4}
\end{figure*}

The cosmic censorship conjecture considered to be true, put forward by Penrose \citep{Penrose:1969pc} has become the pillar of GR. It has being envisaged as a fundamental principle of nature. The strong cosmic censorship conjecture states that such singularities are invisible to all observers. Thus, no NSs for any observer. However, the proof of this conjecture is still an open question and very far from being settled \citep{Joshi:1987wg,Joshi:2000fk} despite the flurry of activity over the years on the conjecture. Given this, it is certainly a worthwhile exercise to investigate photon orbits around NSs. 
The spherical photon orbits around rotating PFDM NSs are depicted in  Figure~\ref{or4}. Clearly, the rotating PFDM NSs have significantly different spherical photon orbits from the Kerr NSs (cf. Figure~\ref{or4} and Table \ref{tab:table2}). 
%%%%%%%%%%%%%%%%%%%%%%%%%%%%%%%%%%%%%%%%%%
%%%%%%%%%%%%%%%%%%%%%%%%%%%%%%%%%%%%%%%%%%%
%%%%%%%%%%%%%%%%%%%%%%%%%%%%%%%%%%%%%%%%%%
%%%%%%%%%%%%%%%%%%%%%%%%%%%%%%%%%%%%%%%%%%%
%%%%%%%%%%%%%%%%%%%%%%%%%%%%%%%%%%%%
\begin{table}
\caption{\label{tab:table2}
Values of various parameters of the photon's trajectory at different $k$ around rotating PFDM NS.}
\centering
%\begin{ruledtabular}
\begin{tabular}{|cccccc|}
\hline
\textbf{$k/M$} & \textbf{$\xi$} & \textbf{$\eta$} & \textbf{$r/M$} & \textbf{$a/M$} & \textbf{$\Delta \phi$} \\ \hline
\multirow{3}{*}{0} & 4.022 & 202.326 & 0.98069 & 1.01 & -37.3006 \\
 & 0.00653 & 20.7745 & 2.1675 & 1.11 & 6.52925 \\
 & -1.44223 & 24.1706 & 1.88956 & $\sqrt{2}$ & -2.33687 \\ \hline
\multirow{3}{*}{0.4} & -0.00598 & 9.58642 & 0.99631 & 1.01 & 6.69319 \\
 & -0.83844 & 11.1342 & 1.79607 & 1.11 & -2.16137 \\
 & -1.87951 & 12.8773 & 1.19303 & $\sqrt{2}$ & 0.54086 \\ \hline
\multirow{3}{*}{0.6} & 0.570361 & 6.19157 & 0.99594 & 1.01 & 12.5605 \\
 & -0.816433 & 10.3119 & 1.88242 & 1.11 & -2.78616 \\
 & -1.18153 & 10.1251 & 1.66013 & $\sqrt{2}$ & -2.00462 \\
 \hline
\end{tabular}
%\end{ruledtabular}
\end{table}
%%%%%%%%%%%%%%%%%%%%%%%%%%%%%%%%%%%%
\subsection{Photon boomerang}
\begin{figure*}[t]
\centering
\includegraphics[width = 0.3\textwidth]{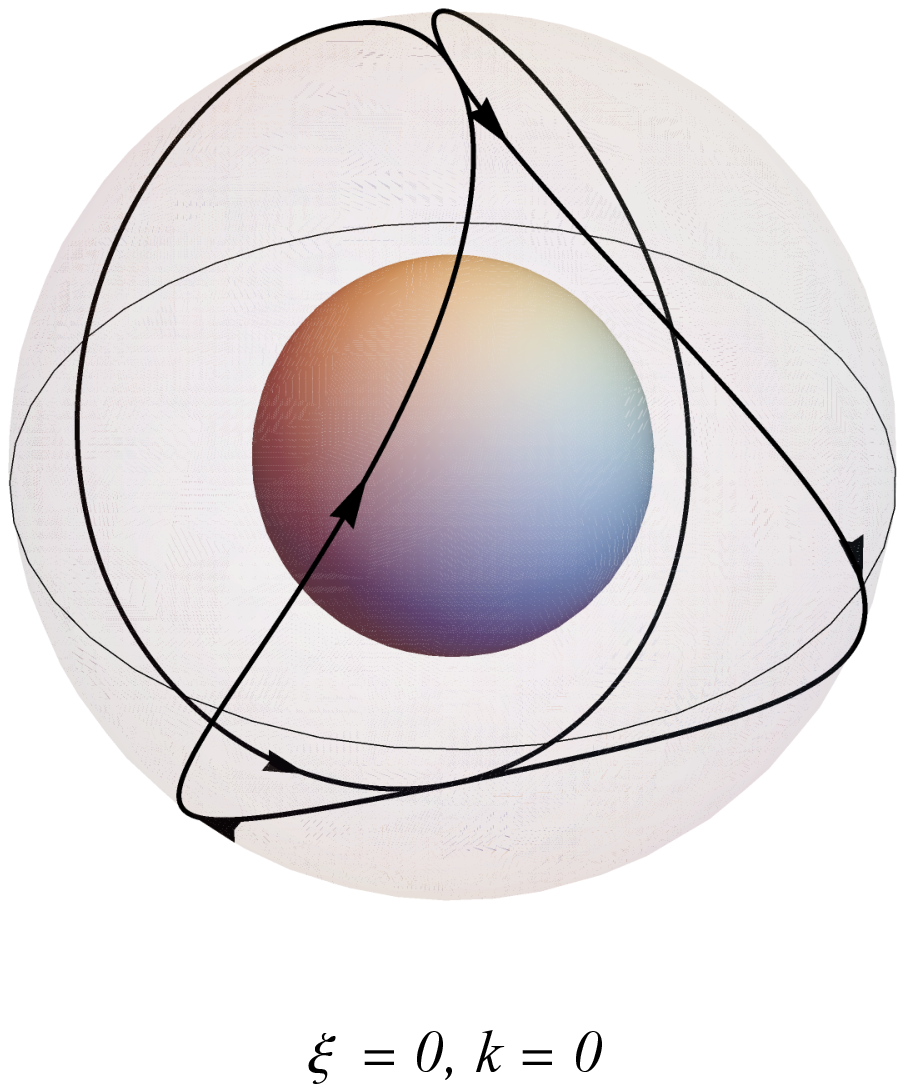} \hspace{0.25cm}
\includegraphics[width = 0.3\textwidth]{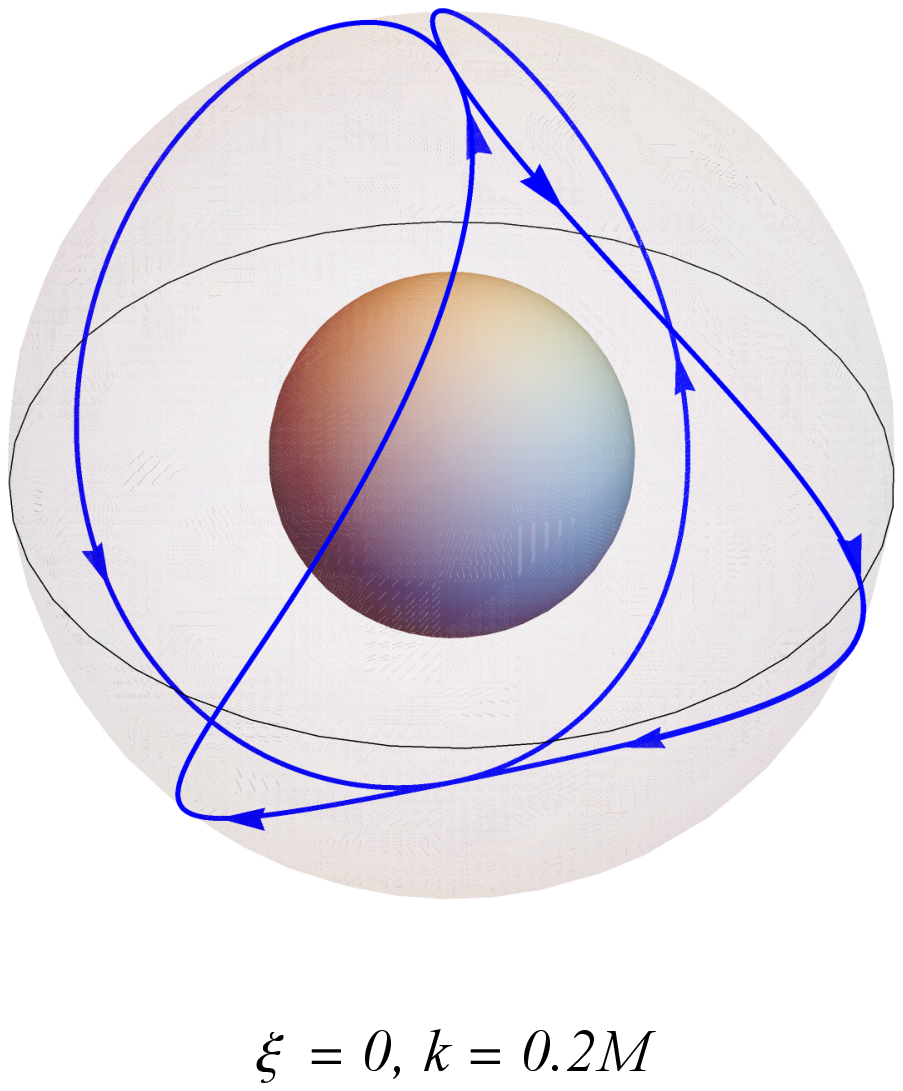} \hspace{0.25cm}
\includegraphics[width = 0.3\textwidth]{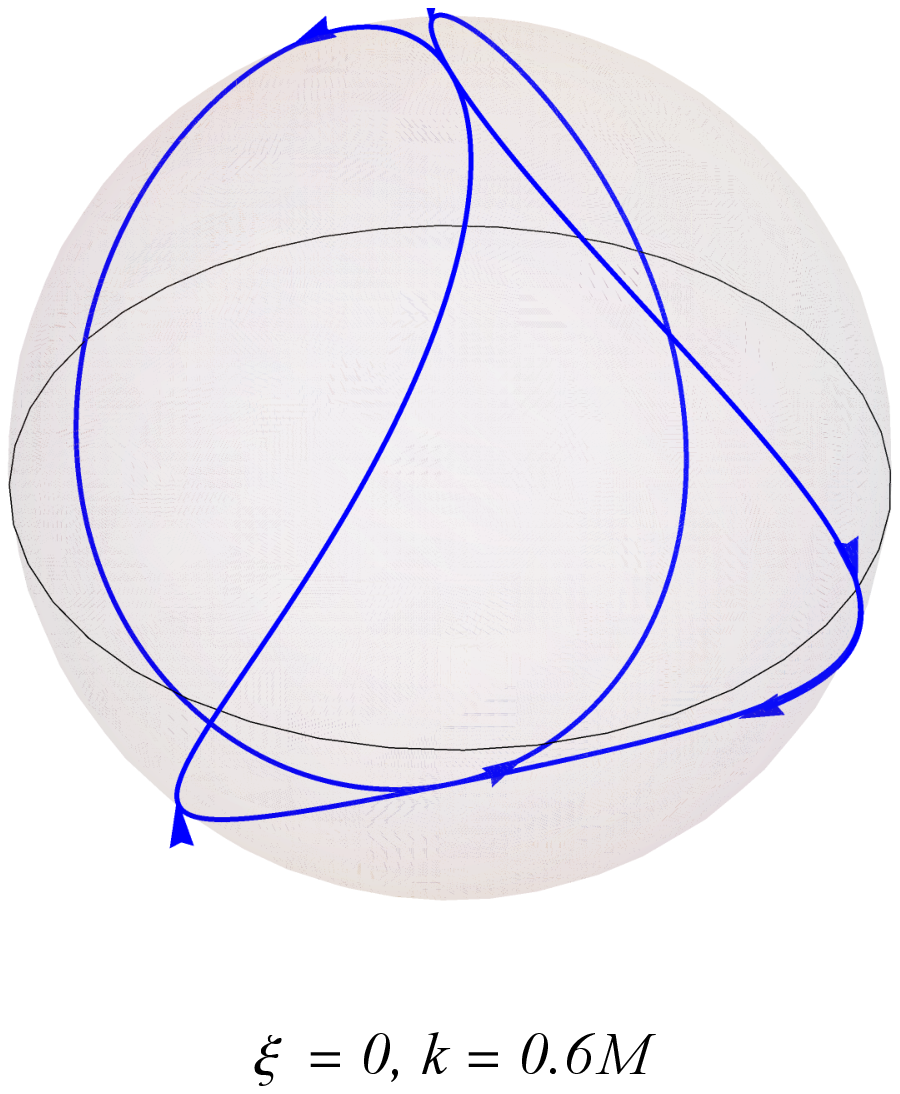}
\caption{Photon \emph{boomerangs} around Kerr BH (left) rotating PFDM BH (middle) and NS (right).} \label{fig:boomerang}
\end{figure*}

When the Kerr BH is near-maximally rotating ($a\approx 0.99434M$), a zero-angular momentum photon shoot perpendicularly from the north polar axis of the counter-rotating BH, at $r\approx 2.4237M$ returns to the same axis but in exactly opposite direction ($\Delta\phi=\pi$) (cf. left panel in Figure~\ref{fig:boomerang}) -- termed photon \emph{boomerang} \citep{Page:2021rhx}. In the Kerr case however, the phenomena of photon \emph{boomerang} cannot occur in case of a NS, wherein, the requirement $a>a_E$ is not fulfilled for any $r$, on solving $\Delta\phi=\pi$. 

Solving Eq. (\ref{tobesolved}) gives the value of $\eta$ and $a$ to produce the \emph{boomerang} orbits. Here, we investigate the effect of the PFDM parameter $k$ on the radius of photon \emph{boomerang}. We solve,
\begin{eqnarray}\label{Delphi_eqn_soln}
\int_{0}^{2\pi} \left( \frac{\frac{a \left(a^2+r^2-a \xi \right)}{\Delta(r)}-\left(a-\frac{\xi }{\sin ^2\theta }\right)}{\sqrt{\eta -\cos ^2\theta \left(\frac{\xi ^2}{\sin ^2 \theta}-a^2\right)}} \right)\,d\theta=\pi
\end{eqnarray}
%%%%%%%%%%%%%%%%%%%%%%%%%%%%%%%%%%%%%%%%%%%%%%%%%%%%%%%%%%%%%%%%%
%%%%%%%%%%%%%%%%%%%%%%%%%%%%%%%%%%%%%%%%%%%%%%%%%%%%%%%%%%%%%%%%%%%
and the numerical integration of Eq.~(\ref{Delphi_eqn_soln}) gives the radius $r_b$ at which the photon \emph{boomerang} happens -- the \emph{boomerang} radius. Photon \emph{boomerangs} for certain values of $a$ and $k$ have been shown in the Figure~ \ref{fig:boomerang}. For the rotating PFDM spacetime, it can be verified from Table~\ref{tab:table5}, that the calculated spin parameter $a$ for photon \emph{boomerangs} may lie in the NS region, this concludes that photon \emph{boomerang} are possible for certain values of $a$ and $k$. On contrary, photon \emph{boomerang} cannot occur for Kerr NSs. The effect of PFDM on parameters associated with photon \emph{boomerang} are shown in Table \ref{tab:table5}. It is evident that $k$ influences the \emph{boomerang} radius $r_b$ as well as the values of $\eta$ at which they occur. We calculate the \emph{boomerang} radii $r_b^{BH}$ and $r_b^{NS}$, modelling supermassive BHs at the centre of different galaxies, respectively, as rotating PFDM BHs and NSs (cf. Table~\ref{tab:table6}).
%%%%%%%%%%%%%%%%%%%%%%%%%%%%%%%%%%%%%%%%%%%%%%%%%%%%%%%%%%%%%%%%%%
%%%%%%%%%%%%%%%%%%%%%%%%%%%%%%%%%%%%%%%%%%%%%%%%%%%%%%%%%%%%%%%%%%%%
\begin{table}
\caption{\label{tab:table5}
Values of various parameters for photon \emph{boomerang} ($\xi$ = 0).}
\centering
%\begin{ruledtabular}
\begin{tabular}{|c|ccccc|}
\hline
\textbf{Spacetime} & \textbf{$k/M$} & \textbf{$\eta$} & \textbf{$r_b/M$} & \textbf{$a/M$} & \textbf{$\Delta \phi$} \\ \hline
\multirow{3}{*}{\begin{tabular}[c]{@{}c@{}} BH\end{tabular}} & 0 & 23.368 & 2.42378 & 0.9943 & 3.14159 \\
 & 0.1 & 15.705 & 2.06692 & 0.8879 & 3.14159 \\
 & 0.2 & 12.728 & 1.89699 & 0.8595 & 3.14159 \\ \hline
\multirow{4}{*}{\begin{tabular}[c]{@{}c@{}} NS\end{tabular}} & 0.3 & 11.004 & 1.79743 & 0.8618 & 3.14159 \\
 & 0.4 & 9.9533 & 1.73941 & 0.8830 & 3.14159 \\
 & 0.5 & 9.3096 & 1.70837 & 0.9172 & 3.14159 \\
 & 0.6 & 8.9311 & 1.69576 & 0.9609 & 3.14159\\
 \hline
\end{tabular}
%\end{ruledtabular}
\end{table}
%%%%%%%%%%%%%%%%%%%%%%%%%%%%%%%%%%%%%%%%%%%%%%%%%%%%%%%%%%%%%%%%%%%%%%%

\begin{table*}
\caption{Spherical photon orbit and \emph{boomerang} radii around supermassive compact objects at galactic centres.}
\centering
\begin{tabular}{|ccccccc|}
\hline
Galaxy & $M( M_{\odot})$ & \begin{tabular}[c]{@{}l@{}}$r_p^{-}$ (in metres)\\ ($k = 0.2M$,\\ $a = 0.8595M$)\end{tabular} & \begin{tabular}[c]{@{}l@{}}$r_p^{*}$ (in metres)\\ ($k = 0.2M$,\\ $a = 0.8595M$)\end{tabular} & \begin{tabular}[c]{@{}l@{}}$r_p^{+}$ (in metres)\\ ($k = 0.2M$,\\ $a = 0.8595M$)\end{tabular} & \begin{tabular}[c]{@{}l@{}}$r_b^{BH}$ (in metres)\\ \,\,($k = 0.2M$,\\ \,\,$a = 0.8595M$,\\ \,\,$\eta = 12.7282$)\end{tabular} & \begin{tabular}[c]{@{}l@{}}\textbf{$r_b^{NS}$} (in metres)\\ \,\,($k = 0.6M$,\\ \,\,$a = 0.9609M$,\\ \,\,$\eta = 8.9311$)\end{tabular} \\ \hline 
$\text{Milky Way}$ & $4\times 10^6$  & $4.6917\times 10^9 $ & $1.1186\times 10^{10} $ & $1.8584\times 10^{10} $ & $7.5049\times 10^{10} $ & $5.2660\times 10^{10}$ \\
$ \text{M87} $ &  $6.5\times 10^9 $   & $7.6200\times 10^{12} $ & $1.8177\times 10^{13} $ & $1.2195\times 10^{14} $ & $3.0199\times 10^{13} $ & $8.5572\times 10^{13}$ \\
$ \text{NGC 4472} $ &  $2.54\times 10^9 $ & $2.9792\times 10^{12} $ & $7.1030\times 10^{12} $ & $4.7650\times 10^{13} $ & $1.1801\times 10^{13} $ & $3.3439\times 10^{13}$ \\
$ \text{NGC 1332} $ &  $1.47\times 10^9 $  & $1.7242\times 10^{12} $ & $4.1108\times 10^{12} $ & $2.7580\times 10^{13} $ & $6.8297\times 10^{12} $ & $1.9352\times 10^{13}$ \\
 $ \text{NGC 4374} $ & $9.25\times 10^8 $  & $1.0849\times 10^{12} $ & $2.5867\times 10^{12} $ & $4.2976\times 10^{12} $ & $1.7355\times 10^{13} $ & $1.2177\times 10^{13}$ \\
$ \text{NGC 1399} $ & $8.81\times 10^8 $  & $1.0333\times 10^{12} $ & $2.4637\times 10^{12} $ & $4.0932\times 10^{12} $ & $1.6529\times 10^{13} $ & $1.1598\times 10^{13}$ \\
 $ \text{NGC 3379} $ & $4.16\times 10^8 $ &   $4.8794\times 10^{11} $ & $1.1633\times 10^{12} $ & $1.9327\times 10^{12} $ & $7.8051\times 10^{12} $ & $5.4760\times 10^{12}$ \\
 $ \text{NGC 4486B} $ & $6\times 10^8 $ &   $7.0375\times 10^{11} $ & $1.6779\times 10^{12} $ & $2.7876\times 10^{12} $ & $1.1257\times 10^{13} $ & $7.8990\times 10^{12}$ \\
$ \text{NGC 1374} $ &  $5.9\times 10^8 $ &   $6.9202\times 10^{11} $ & $1.6499\times 10^{12} $ & $2.7411\times 10^{12} $ & $1.1069\times 10^{13} $ & $7.7674\times 10^{12}$ \\
$ \text{NGC 464} $ & $4.72\times 10^9 $ &   $5.5362\times 10^{12} $ & $1.3199\times 10^{13} $ & $2.1929\times 10^{13} $ & $8.8557\times 10^{13} $ & $6.2139\times 10^{13}$ \\
 $ \text{NGC 3608} $ & $4.65\times 10^8 $ &   $5.4541\times 10^{11} $ & $1.3003\times 10^{12} $ & $2.1604\times 10^{12} $ & $8.7244\times 10^{12} $ & $6.1217\times 10^{12}$ \\
$ \text{NGC 3377} $ &  $1.78\times 10^8 $ &   $2.0878\times 10^{11} $ & $4.9777\times 10^{11} $ & $8.2699\times 10^{11} $ & $3.3396\times 10^{12} $ & $2.3434\times 10^{12}$ \\
$ \text{NGC 4697} $ & $2.02\times 10^8 $ &   $2.3693\times 10^{11} $ & $5.6489\times 10^{11} $ & $9.3850\times 10^{11} $ & $3.7899\times 10^{12} $ & $2.6593\times 10^{12}$ \\
 $ \text{NGC 5128} $ & $5.69\times 10^7 $ &   $6.6739\times 10^{10} $ & $1.5912\times 10^{11} $ & $2.6436\times 10^{11} $ & $1.0676\times 10^{12} $ & $7.4909\times 10^{11}$ \\
$ \text{NGC 1316} $ & $1.69\times 10^8 $ &   $1.9822\times 10^{11} $ & $4.7260\times 10^{11} $ & $7.8518\times 10^{11} $ & $3.1708\times 10^{12} $ & $2.2249\times 10^{12}$ \\
 $ \text{NGC 3607} $ & $1.37\times 10^8 $ &  $1.6069\times 10^{11} $ & $3.8312\times 10^{11} $ & $6.3651\times 10^{11} $ & $2.5704\times 10^{12} $ & $1.8036\times 10^{12}$ \\
$ \text{NGC 4473} $ & $9\times 10^7 $ &   $1.0556\times 10^{11} $ & $2.5168\times 10^{11} $ & $4.1814\times 10^{11} $ & $1.6880\times 10^{12} $ & $1.1848\times 10^{12} $\\
$ \text{NGC 4459} $ & $6.96\times 10^7 $ &   $8.1635\times 10^{10} $ & $1.9463\times 10^{11} $ & $3.2336\times 10^{11} $ & $1.3058\times 10^{12} $ & $9.1629\times 10^{11} $\\
$ \text{M32} $ & $2.45\times 10^6 $ &   $2.8736\times 10^9 $ & $6.8514\times 10^9 $ & $1.1383\times 10^{10} $ & $4.5967\times 10^{10} $ & $3.2254\times 10^{10}$ \\
$ \text{NGC 4486A} $ & $1.44\times 10^7 $ &   $1.6890\times 10^{10} $ & $4.0269\times 10^{10} $ & $6.6903\times 10^{10} $ & $2.7018\times 10^{11} $ & $1.8958\times 10^{11} $\\
$ \text{NGC 4382} $ & $1.3\times 10^7 $ &   $1.5248\times 10^{10} $ & $3.6354\times 10^{10} $ & $6.0398\times 10^{10} $ & $2.4391\times 10^{11} $ & $1.7115\times 10^{11}$ \\
$ \text{CYGNUS A} $ & $2.66\times 10^9 $ &   $3.1199\times 10^{12} $ & $7.4386\times 10^{12} $ & $1.2358\times 10^{13} $ & $4.9907\times 10^{13} $ & $3.5019\times 10^{13}$ \\
\hline
\end{tabular}
%\end{ruledtabular}
\label{tab:table6}
\end{table*}
%%%%%%%%%%%%%%%%%%%%%%     Shadow      %%%%%%%%%%%%%%%%%%%%%%%%%%%%
 \section{Constraining dark matter with EHT results} 
 \begin{figure*}[t]
\centering
\begin{tabular}{c c}
    \hspace{-1cm}\includegraphics[scale=0.7]{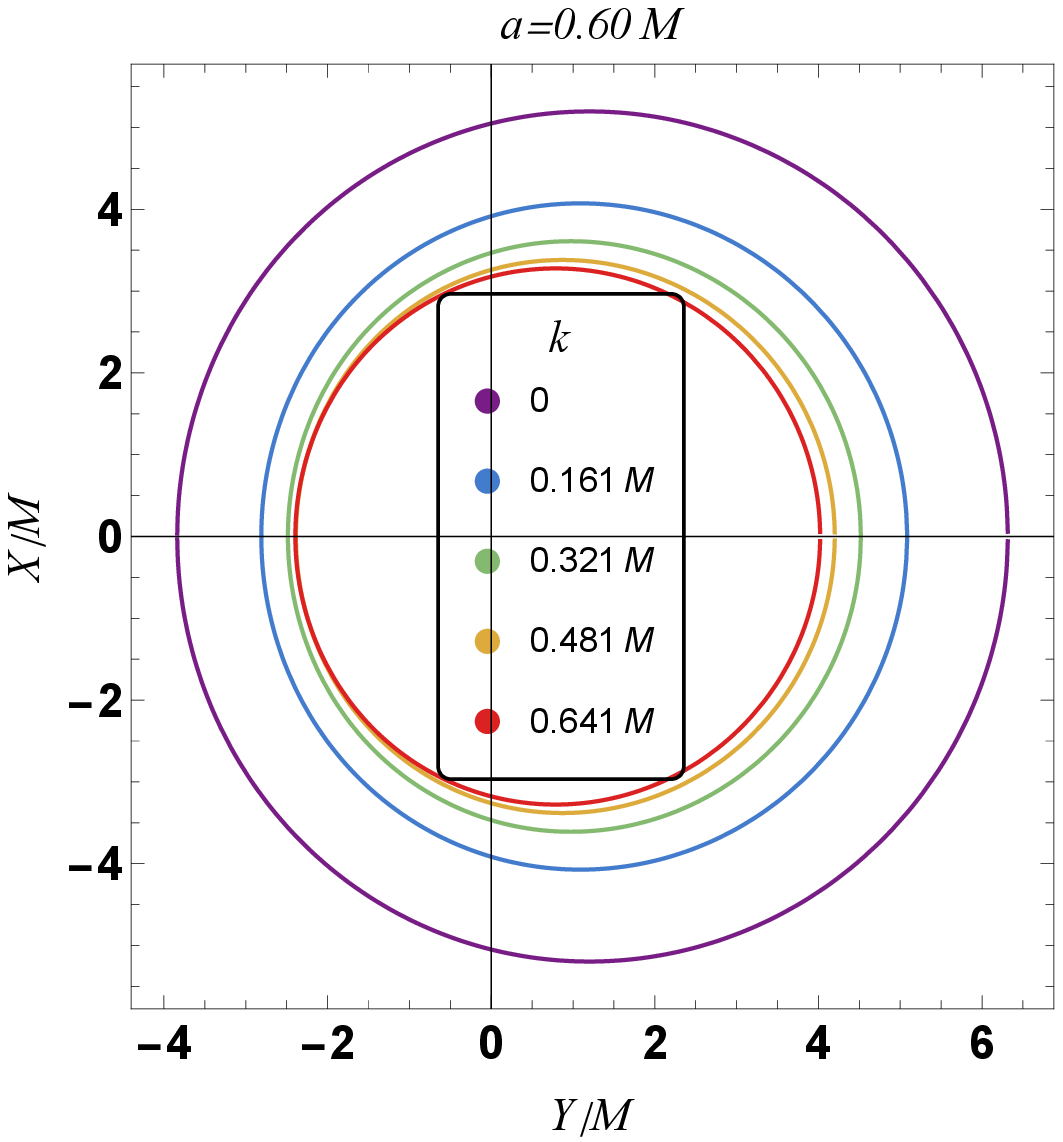}&
    \includegraphics[scale=0.67]{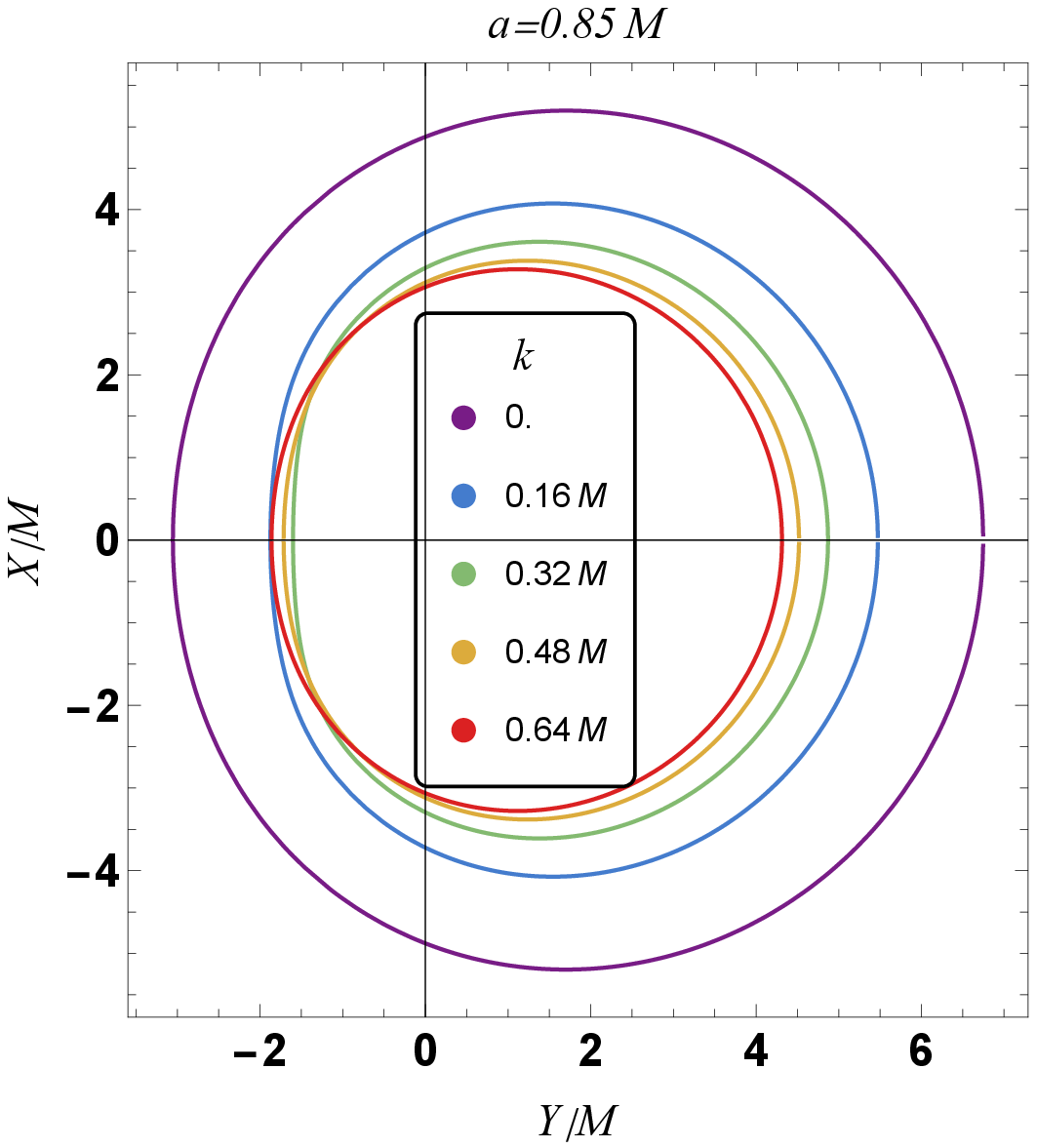}
\end{tabular}
\caption{Shadow cast by the rotating PFDM BHs, with varying parameter $k$ at $\theta_o=90$\textdegree. The violet curves correspond to Kerr BH ($k\to0$) shadows. (For interpretation of the references to colour in this figure legend, the reader is referred to the web version of this article).}\label{fig:shadow}
\end{figure*}\label{sect:constraints}
 %%%%%%%%%%%%%%%%%%%%%%%%%%%%%%%%%%%%%%%%%%%%%%%%%%%%%%%%
The most direct observable manifestation of the null spherical orbital motion, in the vicinity of a BH, is the formation of shadow \citep{Synge:1966okc,Luminet:1979nyg,Bardeen:1973tla} -- a two dimensional dark region in the observer's sky on a bright background, outlined by a series of asymptotically spaced ring like structures that mark the shadow boundary \citep{Johnson:2019ljv,Broderick:2022tfu}. BH shadows in both GR \citep{Falcke:1999pj,Vries2000TheAS,Yumoto:2012kz,Cunha:2018acu,Kumar:2018ple,Afrin:2021ggx,Chael:2021rjo} and MoGs \citep{Amarilla:2010zq,Johannsen:2010gy,Amir:2017slq,Singh:2017vfr,Kumar:2017tdw,Mizuno:2018lxz,Allahyari:2019jqz,Papnoi:2014aaa,Kumar:2019ohr,Kumar:2020bqf,Kumar:2020hgm,Kumar:2020owy,Brahma:2020eos,Ghosh:2020spb,Afrin:2021wlj,Vagnozzi:2022moj,Vagnozzi:2019apd,Afrin:2021imp,KumarWalia:2022aop,Kumar:2022fqo,Islam:2022ybr,Sengo:2022jif,Kuang:2022ojj,Junior:2021svb}, have served as a widely used tool to estimate the parameters associated with the BHs \citep{Hioki:2009na,Kumar:2018ple,Afrin:2021imp,Afrin:2021wlj}, besides, the shadow shape and size can probe the various MoGs and constrain them \citep{Johannsen:2015hib,Bambi:2019tjh,Atamurotov:2013sca,Kumar:2020hgm,Kumar:2020yem,Afrin:2021imp,Afrin:2021wlj,Vagnozzi:2022moj,Afrin:2022ztr}. For constructing shadow, the structure of photon region is necessary, which is composed of spherical photon orbits (as visualized in Figures~\ref{fig:SPO1}-\ref{fig:SPO3}) for both BHs and NSs; we are, however, only interested in the photon region around BHs for studying the shadow formation. The bound spherical photon orbits are unstable, i.e., at $r=r_p\in(r_p^-, r_p^+)$, the ${\mathcal{R}}''\leq0$
and a slight perturbation will result in an exponential divergence -- either towards or away from the BH -- the latter may eventually reach the observer. The ring line structure, namely, the photon ring, is traced by the projection of such nearly bound geodesics that hit the celestial plane of an observer \citep{Johnson:2019ljv}.

The shadow shape and size -- which encode in them information regarding the background theory of gravity -- are dependent on intrinsic parameters like BH mass $M$, spin $a$ as well as other deviation parameters or hairs \citep{Johannsen:2015mdd,Afrin:2021imp,Afrin:2021ggx} which in the present scenario would be the parameter $k$; besides, the shadow characteristics are also dependent on extrinsic parameters like the radial distance of the observer $r_o$ and the inclination angle $\theta_o$. However, for observations made by the EHT, we can practically set off $r_o\to\infty$ and one can visualize the BH shadow outlined by the celestial coordinates defined by \citep{Bardeen:1973tla,Afrin:2021imp}
\begin{equation}
\{X,Y\}=\{-\xi_c\csc\theta_o,\, \pm\sqrt{\eta_c+a^2\cos^2\theta_o-\xi^2\cot^2\theta_o}\}\,\label{Celestial1}
\end{equation}
where the critical impact parameters are given by Eqs.~(\ref{sol2a}) and (\ref{sol2b}). A parametric plot of $(X, Y)$ as a function of $r=r_p\in(r_p^-, r_p^+)$ yields the shadow boundary as shown in Figure~\ref{fig:shadow}. The shadow structure shows interesting behaviour with the onset of dark matter imprints -- the size becomes monotonically smaller and the distortion first increases and then decreases with increase in $k$ which can be directly correlated to similar behaviour of the spherical photon orbit radii (cf. Table~\ref{Table:r1r2_1} and Figure~\ref{fig:u}). Furthermore there is a shift in the shadow centre first towards right end with $k$, which is similar to the effect of increasing $a$ in the Kerr case, owing to Lens-Thirring effect; but one distinguishing feature is the shifting of the shadow centre towards the left end, with further increment in $k$ after a certain value, which is not observed for Kerr BHs.
Interestingly, the impact of parameter $k$ the shadow deformation is analogous to that of the $a$ on the Kerr BH shadow due to which there is a possibility of degeneracy in the shadow shapes of the rotating PFDM BHs with some parameter $k$ and $a$ and that cast by the Kerr BH with some spin $a^*$. We will exploit this possible degeneracy in the shadow characteristics of the rotating PFDM BH and the Kerr BH to constrain dark matter imprints.

To astrophysically constrain the parameter $k$, we introduce the shadow observable -- Schawarzschild deviation $\delta$, for which we must first numerically obtain the shadow area as \citep{Abdujabbarov:2015xqa,Kumar:2018ple},
\begin{eqnarray}
A&=&2\int{Y(r_p) dX(r_p)}\nonumber\\
&=&2\int_{r_p^{-}}^{r_p^+}\left( Y(r_p) \frac{dX(r_p)}{dr_p}\right)dr_p,\label{Area}
\end{eqnarray} 
%%%%%%%%%%%%%%%%%%%%%%%%%%%%%%%%%%%%%%%%%%%%%%%%%%%%%%%%%%%%%%%%%%%%%%
\begin{figure*}[t]
\begin{center}
    \begin{tabular}{c c}
    \hspace{-1cm} \includegraphics[scale=0.8]{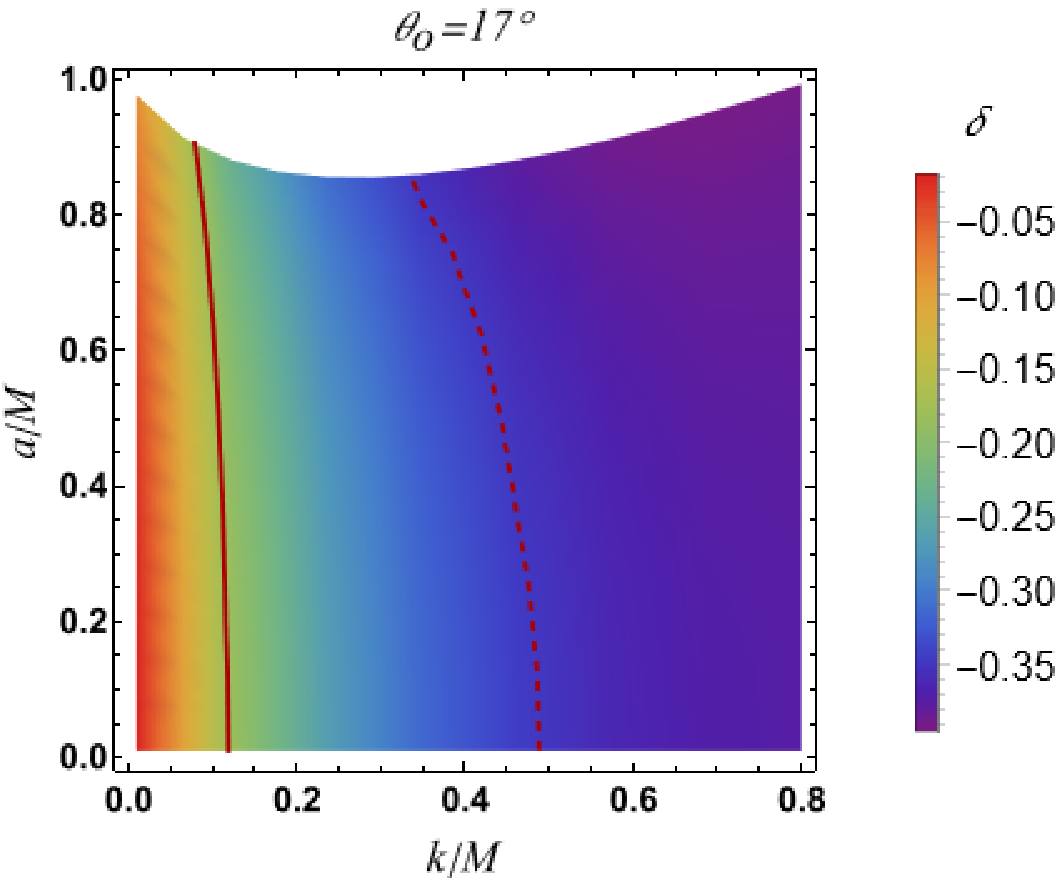}&
     \includegraphics[scale=0.8]{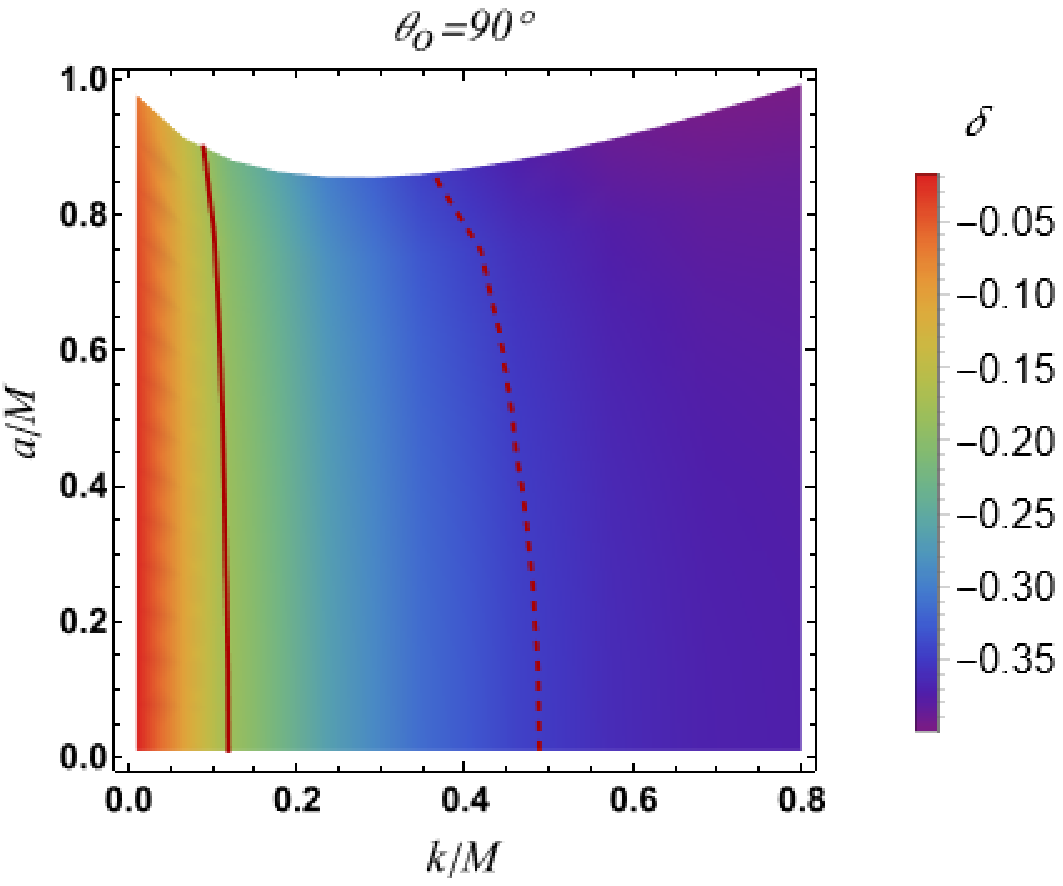}
\end{tabular}
\end{center}
	\caption{Constraints from EHT results of Schwarzschild shadow deviation  $\delta$, modelling M87* as rotating PFDM BH. The solid and dashed curves correspond, respectively, to the $1\sigma$ and $2\sigma$ bounds of the measured Schwarzschild deviation, $\delta=-0.01\pm0.17$ of M87*, as reported by the EHT observations.}
	\label{Fig:shadowDiameter_M87}
\end{figure*}
%%%%%%%%%%%%%%%%%%%%%%%%%%%%%%%%%%%%%%%%%%%%%%%%%%%%
\begin{figure*}[t]
\begin{center}
    \begin{tabular}{c c}
    \hspace{-1cm} \includegraphics[scale=0.8]{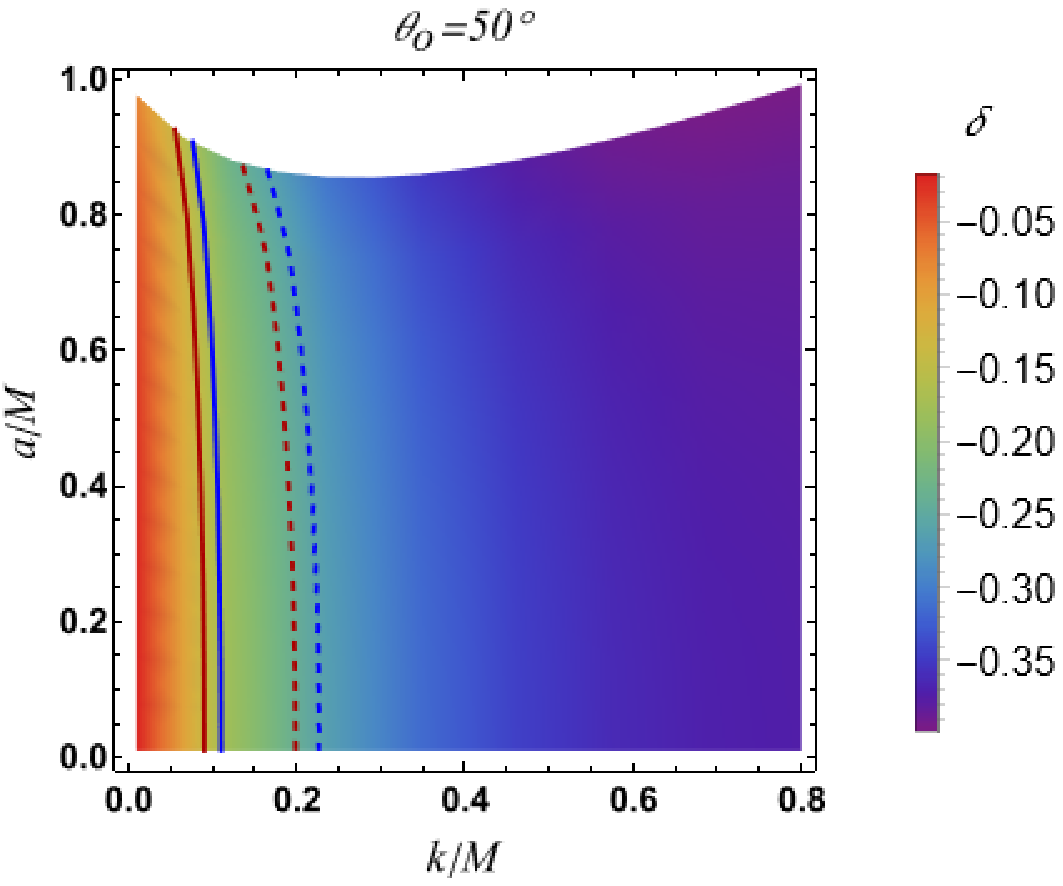}&
     \includegraphics[scale=0.8]{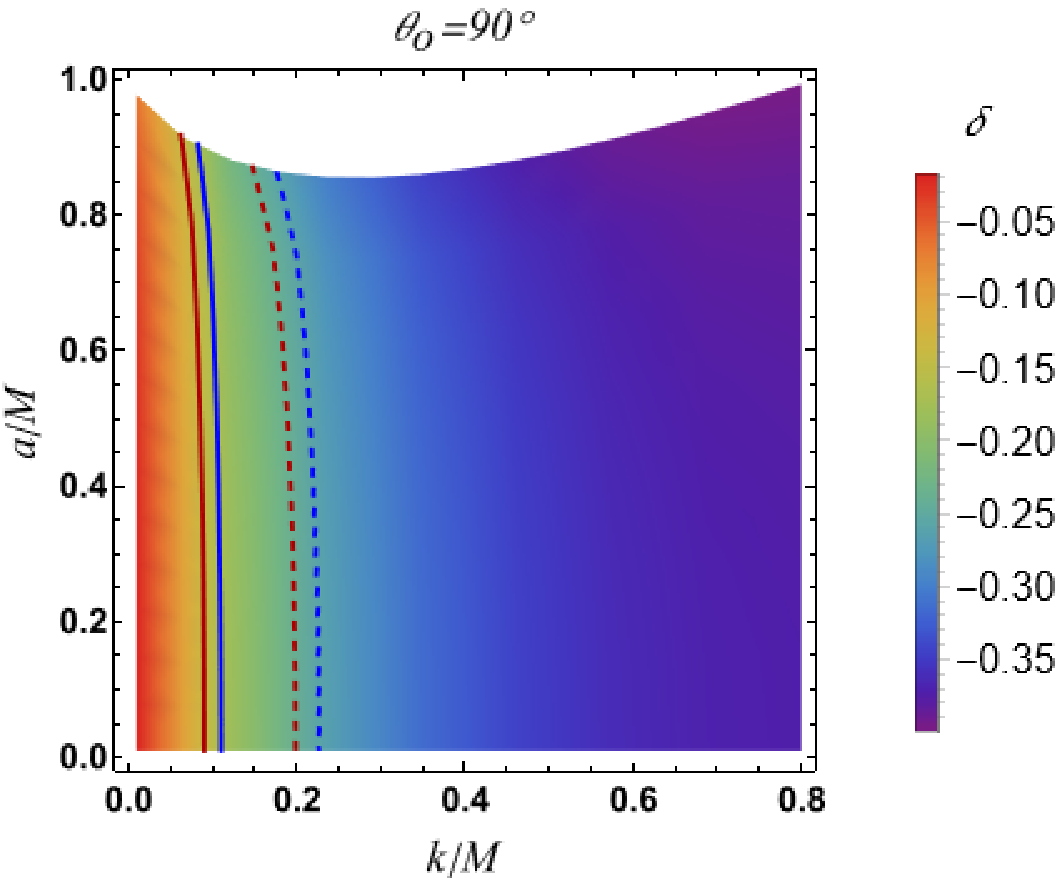}
\end{tabular}
\end{center}
	\caption{Constraints from EHT results of Schwarzschild shadow deviation  $\delta$, modelling Sgr A* as rotating PFDM BH. The blue and red solid contours correspond, respectively, to the $1\sigma$ bounds of the measured Schwarzschild deviation $\delta= -0.08^{+0.09}_{-0.09}~\text{(VLTI)},-0.04^{+0.09}_{-0.10}~\text{(Keck)}$ of Sgr A*, as reported by the EHT observations. The dashed lines correspond to the respective $2\sigma$ bounds. (For interpretation of the references to colour in this figure legend, the reader is referred to the web version of this article).}
	\label{Fig:shadowDiameter_SgrA}
\end{figure*}
%%%%%%%%%%%%%%%%%%%%%%%%%%%%%%%%%%%%%%%%%%%%%%%%%%%%%%%%%%%%%%
\begin{table*}[tbh]
\centering
\begin{tabular}{|c|cc|}
\hline
\multirow{2}{*}{\begin{tabular}[c]{@{}c@{}}Shadow\\ observable\end{tabular}} & \multicolumn{2}{c|}{Constraints}                                                     \\ \cline{2-3} 
                 & \multicolumn{1}{c|}{$1\sigma$}           & $2\sigma$           \\ \hline
$\delta_{M87^*}$ & \multicolumn{1}{c|}{$k\in [0, 0.0792M)$} & $k\in [0, 0.3349M)$ \\ \hline
$\delta_{Sgr A^*}$                                                           & \multicolumn{1}{c|}{$k^{max}\in(0.0507M, 0.0611M)$} & $k^{max}\in(0.1282M, 0.1489M)$ \\ \hline
\end{tabular}
\caption{Constraints on the PFDM parameter $k$ from EHT results of M87* and Sgr A* black holes.}
\label{Table:constraints}
\end{table*}
%%%%%%%%%%%%%%%%%%%%%%%%%%%%%%%%%%%%%%%%%%%%%%%%
Modelling the supermassive BHs M87* and Sgr A* as rotating PFDM BHs, the Schwarzschild shadow deviation ($\delta$) would quantify the difference between the model shadow diameter ($\Tilde{d}_{metric}$) and the Schwarzschild shadow diameter $6\sqrt{3}M$ as \citep{EventHorizonTelescope:2022xnr,EventHorizonTelescope:2022xqj},
\begin{equation}\label{SchwarzschildShadowDiameter}
    \delta=\frac{\Tilde{d}_{metric}}{6\sqrt{3}}-1.
\end{equation}
Here $\Tilde{d}_{metric}=2R_a$ is the model shadow diameter with $R_a=\sqrt{A/\pi}$. The EHT has recently released  images of both the M87* and Sgr A* which exhibit a luminous bright ring of emission, with diameters 42 $\pm$ 3 $\mu$as and 51.8 $\pm$ 2.3 $\mu$as (68\% credible interval) respectively, surrounding a central brightness depression -- which besides meeting the expectations of the presence of a supermassive BH \citep{EventHorizonTelescope:2019dse,EventHorizonTelescope:2022xnr}, confirm accord with GR  \citep{EventHorizonTelescope:2019dse,EventHorizonTelescope:2022xqj,EventHorizonTelescope:2022xnr,EventHorizonTelescope:2022xqj}. However, the M87* observations disfavour NSs \citep{EventHorizonTelescope:2019dse,EventHorizonTelescope:2019pgp,EventHorizonTelescope:2019ggy}, hence we attempt to analyse only BH case here.
The EHT observational results put bounds on $\delta$ of the dark region, which can be identified as the BH shadow, for the M87* and Sgr A* BHs, besides inferring their masses, $M_{M87^*}= 6.5\pm0.2\times 10^9 M_\odot$ and $M_{SgrA^*} = 4.0 \times 10^6 M_\odot $ \citep{EventHorizonTelescope:2019dse,EventHorizonTelescope:2019pgp,EventHorizonTelescope:2022xnr,EventHorizonTelescope:2022xqj}, and distances $d_{M87^*}=16.8^{+0.5}_{-0.7}Mpc$ and $d_{SgrA^*} =$ \citep{EventHorizonTelescope:2019dse,EventHorizonTelescope:2019pgp,EventHorizonTelescope:2022xnr,EventHorizonTelescope:2022xqj} from earth. We aim to utilize these inferred bounds to constraint the deviation of the rotating PFDM BHs in question from Kerr BHs. However, we note that, the possible uncertainties in inferred mass and distance of the two target BHs are already taken into account by the EHT and get reflected in the resulting bounds; for the sake of simplicity, we do not consider the mass and distance uncertainties in our analysis and the method would yield constraints on the BH parameters in exactly the same way for different mass and distance priors as well. However,  one of the major sources of uncertainty in the EHT results comes from that of the in-determination of $\theta_o$, is tackled by considering the full range $\theta_o\in[0$\textdegree$, 90$\textdegree$]$.
\paragraph*{Constraints from M87* results--}
In 2019, the EHT collaboration had released the first ever image of the supermassive BH M87* and using an extensive library of stimulated BH images, has resolved the central compact radio source to be an asymmetric bright emission ring with a diameter of $42\pm3\mu$as \citep{EventHorizonTelescope:2019dse,EventHorizonTelescope:2019pgp}, which is calibrated with the shadow size of M87*, to obtain the $\delta_{M87^*}=-0.01\pm0.17$ within $1\sigma$ confidence region \citep{EventHorizonTelescope:2019ggy,EventHorizonTelescope:2021dqv}. Considering the inferred BH mass $M_{M87^*}= 6.5\times10^9 M_\odot$ and distance $d_{M87^*}=16.8 Mpc$, we impose the bound on $\delta_{M87^*}$ (cf. Figure~\ref{Fig:shadowDiameter_M87}) to obtain the following constraints: $k\in [0, 0.3607M)$ 
within $2\sigma$, $k\in [0, 0.0895M)$ within $1\sigma$ confidence levels at $\theta_0=90$\textdegree and $k\in [0, 0.3349M)$ 
within $2\sigma$, $k\in [0, 0.0792M)$ within $1\sigma$ confidence levels at $\theta_0=17$\textdegree. Thus we infer that for $0 \leq k< 0.0792M$, the M87* can be a rotating PFDM BH at the current precision of the astrophysical observations.
\paragraph*{Constraints from Sgr A* results --}
Though the results of M87* give a scope to constrain dark matter, yet, the Sgr A* would give an independent test at a curvature scale $\sim\mathcal{O}(10^3)$ higher than that offered by M87*, that would give complementary results. Further, the independent measurement of mass-to-distance ratio of Sgr A* through stellar astrometry via Keck telescopes and Very Large Telescope Interferometer (VLTI) \citep{Do:2019txf,GRAVITY2019,GRAVITY2021,GRAVITY2022,EventHorizonTelescope:2022xqj} give a better and parameter-free predictions of spacetime properties from the captured image of Sgr A* \citep{EventHorizonTelescope:2022xqj}. We would thus utilize the EHT results for Sgr A* now, to put bounds on $k$. We utilize the mass $M_{SgrA^*} = 4.0 \times 10^6 M_\odot $ and distance $d_{SgrA^*}=8 kpc$ priors \citep{EventHorizonTelescope:2022xnr,EventHorizonTelescope:2022xqj} along with the bound on the shadow observable -- $\delta_{Sgr A^*} = -0.08^{+0.09}_{-0.09}~\text{(VLTI)},-0.04^{+0.09}_{-0.10}~\text{(Keck)}$ at $1\sigma$ confidence level \citep{EventHorizonTelescope:2022xnr,EventHorizonTelescope:2022xqj} -- to find the accordant ranges of the parameter $k$ as: $k\in [0, 0.1489M)$ 
within $2\sigma$, $k\in [0, 0.0611M)$ within $1\sigma$ confidence levels at $\theta_0=90$\textdegree, $k\in [0, 0.1386M)$ 
within $2\sigma$, $k\in [0, 0.0559M)$ within $1\sigma$  confidence levels at $\theta_0=50$\textdegree and $k\in [0, 0.1282M)$ 
within $2\sigma$, $k\in [0, 0.0507M)$ within $1\sigma$ confidence levels at $\theta_0=0$\textdegree \, (cf. Figure~\ref{Fig:shadowDiameter_SgrA}). Thus as $\theta_o$ varies from $0$\textdegree to $90$\textdegree, the upper bound on $k$ varies as $k^{max}\in(0.0507M, 0.0611M)$. The results of Sgr A* puts more stringent constraints than those of M87*. We thus report, as a first, the possible range of $k$ spanning over the entire range of inclinations. We tabulate the astrophysical constraints on $k$ in Table~\ref{Table:constraints}.
\section{Conclusion} \label{Conclusion}
For various astrophysical phenomena like strong field gravitational lensing and shadow formation, the path of light near the BH is crucial. The radii of the polar and equatorial plane light orbits have previously been determined explicitly in terms of the BH's spin parameter. The present study has expanded the previous analytical solutions of null geodesics in Kerr spacetime with surrounding PFDM. We have shown that rotating PFDM BHs and NSs can have spherical photon orbits around them. We determine the parameter space segregating the rotating PFDM BHs from the NSs and get the parameters for obtaining a particular class of orbits called photon \emph{boomerang} orbits in both the BH and NS cases. The radii of spherical photon orbits $r_p^{\pm}, r_p^*$ along with the event horizon, $r_+$ are found first to decrease and then increase with $k$. In a single latitudinal oscillation of the spherical photon orbits around the rotating PFDM BHs, the azimuthal oscillations vary with $k$ -- $\Delta \phi$ first increases and then decreases in the case of prograde orbits whereas for the retrograde orbits, $\Delta\phi$ first decreases and then increases with $k$. In contrast to the Kerr BH ($k=0$), where the azimuthal oscillations monotonically increase with increasing $\xi$, the $\Delta\phi$ shows no monotonic behaviour with $\xi$ in case of rotating PFDM BHs. We have also analysed the photon \emph{boomerang} around the rotating PFDM BHs and NSs. Unlike the Kerr NSs, photon \emph{boomerangs} can be formed for rotating PFDM NSs. We find the parameter $k$ can influence the \emph{boomerang} orbit radius $r_b$ and the values of $\eta$ at which they occur. We have found the radii of \emph{boomerang} orbits and the event horizon and spherical photon orbit radii near several supermassive black holes considering surrounding PFDM.

We have also analysed the astrophysical implications of the effects of PFDM on spherical photon trajectories by investigating how the PFDM affects the characteristics of shadows cast by the rotating PFDM BHs. Indeed, the shadow size becomes monotonically smaller, and the distortion first increases and then decreases with an increase in $k$; it can directly correlate to similar behaviour exhibited by spherical photon orbit radii. Further, the shadow centre first shifts towards right with $k$, which is like the effect of increasing $a$ in the Kerr BHs; but one distinguishing feature from the Kerr BH is the shifting of the shadow centre towards the left end, with further increment in $k$ after a specific value. These distinguishing shadow features motivated us to explore the possibility of utilizing the deviation of the model shadow size from the Schwarzschild shadow size to put constraints on the PFDM. 
By modelling the rotating PFDM BHs as the supermassive BHs M87* and Sgr A*, using the inferred BH mass and distances, we impose the EHT bounds on the Schwarzschild shadow deviation to conclude, respectively, $0\leq k\leq 0.0792M$ and $k^{max}\in(0.0507M, 0.0611M)$. The results of Sgr A* puts more stringent constraints on $k$ than those of M87*. Thus in a concordant finite parameter space of the rotating PFDM BHs, the EHT observations do not rule out the possibility of surrounding PFDM at galactic centres. In conclusion, The EHT observation has provided the first direct image of a supermassive black hole M87* and Sgr A*. 
We have achieved one of the first constraints on the dark matter with the EHT results of M87* and Sgr A*. It allows for analyzing astrophysical phenomena like strong gravitational lensing and accretion mechanisms around supermassive black holes to get positive results on the traces of galactic dark matter.  Further, the future EHT experiments is expected to resolve the photon sub-rings -- which entails novel observables like universal photon ring autocorrelations \citep{Hadar:2020fda} that have been used to constrain the dark matter \citep{Chen:2022kzv} -- whose structure and location are directly determined by the properties of photon region \citep{Hadar:2020fda}.

Since the present resolution of the EHT images is of $\mathcal{O}(\mu\text{as})$, distinguishing the PFDM and Kerr BHs is difficult and with future more precise ground and space based \citep{Johnson:2019ljv,Andrianov:2022snn} imaging technologies, like the ngEHT \citep{Blackburn:2019bly}, we can pin down the exact constraint on $k$, which, in principle, may provide a signature of dark matter.

\section*{Acknowledgements}
A.A. would like to thank Shafqat ul Islam for his useful insight and help in plots. A.A. is supported by DST-INSPIRE scholarship and M.A.\, is supported by a DST-INSPIRE fellowship, Department of Science and Technology, Government of India. S.G.G. thanks SERB-DST for research project No.~CRG/2021/005771.

%\clearpage 
\bibliography{bibfile}
\end{document}